# Uniaxial spin texture in a superconducting electron gas revealed by exchange interactions

Junyi Yang[1*], Changjiang Liu[1,2*], Xianjing Zhou[3], Hanyu Hou[4], Kaijun Yin[4], Jianguo Wen[3], John Pearson[1,3], Alexey Suslov[5], Dafei Jin[3,6], Jidong S. Jiang[1], Ulrich Welp[1], Jian-Min Zuo[4], Michael R. Norman[1], Anand Bhattacharya[1]

1. Materials Science Division, Argonne National Laboratory, Lemont, USA
2. Department of Physics, University at Buffalo, SUNY, Buffalo, USA
3. Center for Nanoscale Materials, Argonne National Laboratory, Lemont, USA
4. Department of Material Science and Engineering, University of Illinois at Urbana-Champaign, Urbana, USA
5. National High Field Magnetic Laboratory, Tallahassee, USA
6. University of Notre Dame, Notre Dame, USA

## Abstract

Two-dimensional superconductors with spin-textured Fermi surfaces can be a platform for realizing unconventional pairing states and are of substantial interest in the context of quantum information science, and superconducting spintronics/orbitronics. We observed an unusual in-plane uniaxial anisotropy in the superconducting 2D electron gas (2DEG) formed at $EuO_x/KTaO_3$ (110) interfaces. This anisotropy is not evident in $AlO_x/KTaO_3$ (110) where the overlayer is non-magnetic. Our results are consistent with a highly anisotropic 'half-Rashba' spin-textured Fermi surface in 2DEGs formed at the $KTaO_3$ (110) interface that is hidden from external magnetic fields due to a near cancellation between orbital and spin moments but revealed by exchange interactions of the electrons in the 2DEG with Eu moments near the $EuO_x/KTaO_3$ (110) interface. The interactions between the uniaxial spin texture and the magnetic overlayer offer new ways to explore the interplay between magnetism and 2D superconductivity.

*equal contribution

## Introduction

The interplay of large spin-orbit coupling (SOC) with crystalline symmetry breaking can result in non-trivial spin-textures for electrons at the Fermi surface of a two-dimensional electron gas (2DEG) [1, 2] . Broken inversion symmetry along the out-of-plane direction in 2DEGs leads to a 'Rashba' spin texture, where spins are locked in-plane with opposite helicity on spin-split Fermi surfaces [3], and such asymmetric SOC also allows for superconductivity with mixed parity [4, 5]. The presence of strong SOC with in-plane inversion symmetry breaking, or broken orbital degeneracy, can lead to an 'Ising' spin texture that locks spins along the out-of-plane direction, and for superconducting 2DEGs this results in large in-plane critical fields [6-8] well in excess of the Pauli limit and may lead to unconventional superconductivity [9]. These spin textures are nominally not expected to give rise to an anisotropic response to in-plane magnetic fields. Rather, such a response is often associated with effective mass anisotropy, nematicity [10] and superconducting order parameters with mixed-parity [11]. On the other hand, an asymmetric strong SOC in a 2DEG with reduced symmetry can result in an anisotropic Rashba interaction, where the spins may have an in-plane anisotropy, or even be locked-in along an axis for a 'half-Rashba' interaction [12]. Such a uniaxial spin texture at the Fermi surface would lead to superconductivity with inherent in-plane anisotropy, which is of both fundamental interest [13] and may also be relevant for applications in quantum information science [14], superconducting spintronics [15] and orbitronics [16].

Superconductivity was recently discovered in 2DEGs formed at interfaces of $KTaO_3$ (KTO) [17] . Notably, the $T_c$ in KTO 2DEGs was found to strongly depend on the orientation of the interface [17-19] and it was proposed that the orientation dependence of $T_c$ is due to varying levels of degeneracy in the $t_{2g}$ manifold of the Ta-5$d$ bands in quantum-confined 2DEGs formed on different crystalline facets of KTO [19]. For the KTO (110) interface (Fig.1 (a)), the degenerate $d_{xz}/d_{yz}$ states at the $\Gamma$ point are lower in energy than $d_{xy}$ due to confinement along the [110] axis and they form an anisotropic Fermi surface with larger/smaller effective mass along the $[1\bar{1}0]$/[001] axis [20], respectively. Furthermore, the $d_{xz}/d_{yz}$ states at $\Gamma$ form the combination $d_{yz} \pm \mathrm{i}d_{xz}$ with an orbital angular momentum axis oriented along the in-plane [001] direction. Due to SOC, the spins anti-align to these orbital moments, and this gives rise to an in-plane anisotropy for both orbital and spin textures along [001] with no spin canting along the out-of-plane direction. Additionally, broken inversion symmetry at the KTO (110) interface gives rise to Rashba splitting of bands formed from these states, which along with the uniaxial anisotropy induces a dominant $\boldsymbol{k}_y\hat{\sigma}_x$ 'half-Rashba' texture due to spin-momentum locking (here $\boldsymbol{x}$ refers to [001] and $\boldsymbol{y}$ to [1$\bar{1}$0]) (Fig. 1(b)). We note that this uniaxial anisotropy is unique to the KTO (110) interface and distinct from the more isotropic spin textures calculated for KTO (111) [21] and KTO (001) [22] interfacial 2DEGs. Lastly, as is well known, for single electron occupancy in the $t_{2g}$ manifold, the orbital and spin moments are 'anti-aligned', which leads to a strongly reduced $g$-factor [23], and a reduced Zeeman coupling of external magnetic fields [24] to the electronic spin texture. Thus, KTO interfacial 2DEGs are nominally expected to have a relatively weak response to in-plane magnetic fields.

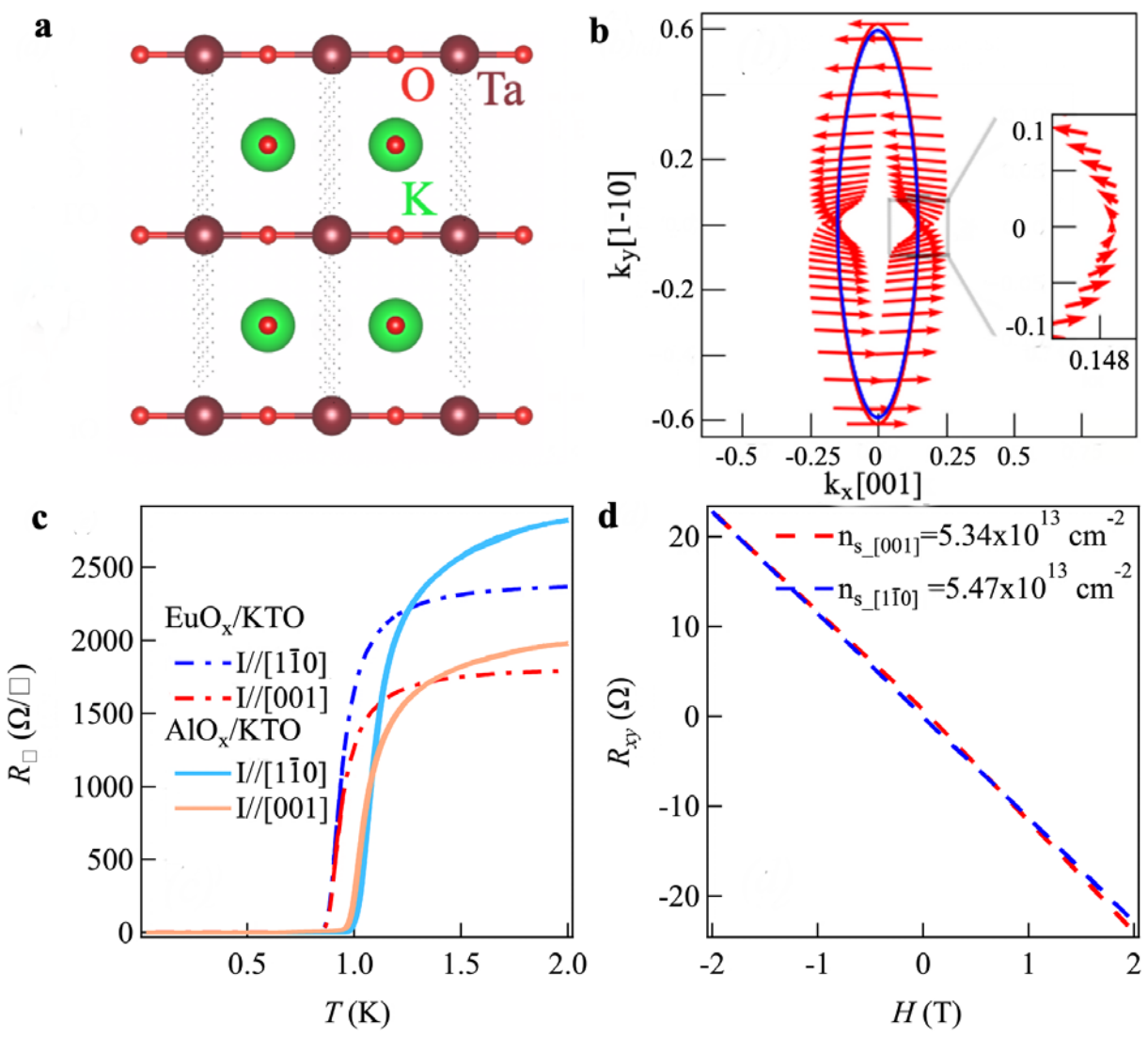

**Figure 1.** (a) Schematic diagram of KTO(110). (b) KTO (110) Fermi surface derived from $d_{xz}/d_{yz}$ states, with its anisotropic spin texture. The length of the arrows is related to the spin vector size, with the inset showing an expanded view around [001]. The two Rashba split surfaces are shown in red and blue, but the spin texture of only one of them is presented for clarity. (c) Temperature dependence of the sheet resistance of $EuO_x$/KTO and $AlO_x$/KTO with current applied along two different in-plane directions. The light blue solid line denotes the current along [1$\bar{1}$0] and the light red solid line denotes the current along [001] for $AlO_x$/KTO. The blue dashed line denotes the current along [1$\bar{1}$0] and the red dashed line denotes the current along [001] for $EuO_x$/KTO. (d) Hall measurements for Hall bar devices along the [001]/ [1$\bar{1}$0] directions, respectively.

## Results

In this work we report on a uniaxial in-plane spin texture for $KTO_3$ (110) interfaces which is revealed by the anisotropy of the in-plane upper critical field $H_c$ as well as the normal state magnetoresistance for $EuO_x$/KTO (110) 2DEGs, where the $EuO_x$ overlayer is magnetic. We fabricated Hall bar devices on both $AlO_x$/KTO (110) and $EuO_x$/KTO (110) (Fig. S1) 2DEGs. Here $AlO_x$ is non-magnetic, while $EuO_x$ is ferromagnetic with $T_{Curie}$ ~ 70 K. Figure 1(c) shows the temperature dependence of the sheet resistance ( $R_{\square}$ ) for $EuO_x$/KTO with current along [001]/[1$\bar{1}$0], respectively. The crystallographic axis for each direction is determined through x-ray diffraction (Fig. S2). Superconductivity is observed for both samples below 800 mK. The normal state $R_{\square}$ for a current along [1$\bar{1}$0] is found to be larger than that for a current along [001] by a factor of 1.5 for all samples. These observations are consistent with a highly anisotropic Fermi surface of KTO (110) [20] (Fig. 1(b)), where the effective mass along the [1$\bar{1}$0] direction is larger than that along [001]. Hall measurements (Fig. 1(d)) along these two crystallographic directions are nearly identical, and presumably the anisotropic $R_{\square}$ is due to the anisotropy in electron mobility. We note that for LAO/STO (110), a similar anisotropy in $R_{\square}$ is observed, though it changes as a function of carrier density [25-27] due to occupation of the $d_{xy}$ bands at higher densities. In contrast, our KTO (110) 2DEGs, the anisotropy of the normal state $R_{\square}$ is found to be similar for both low and high densities. This may imply that the $d_{xy}$ bands are not occupied or have fewer carriers than the $d_{xz}/d_{yz}$ band in our KTO (110) 2DEGs, presumably due to the stronger confinement effects in KTO (110) relative to STO (110). However, we note that recent angle-resolved photoemission experiments on doped KTO (110) surfaces with higher carrier density (~ 7 x $10^{13}$/cm$^2$ ) do observe the occupation of the higher lying $d_{xy}$ bands [20].

**Anisotropic superconductor-ferromagnet interaction in $EuO_x$/KTO (110) near $T_c$**

To explore the anisotropic spin texture in KTO (110), we first measured the in-plane magnetic field dependence of $R_{\square}$ for $EuO_x$/KTO (110). We find evidence for coupling between conduction electrons in the 2DEGs and the magnetic $EuO_x$ overlayer in measurements of $R_{\square}$(H) at 2 K (above $T_c$), where we observe hysteretic behavior at low fields for $H$ // [001] (Fig. 2 (a)). Notably, the MR $= \frac{R(H)-R(0)}{R(0)}$ and apparent hysteresis for $H$ // [001] is much larger than that for $H$ // [1$\bar{1}$0]. The hysteretic minima in $R_{\square}$(H) at ± 115 Oe are consistent with the in-plane coercive field of the $EuO_x$ overlayer of ~ 115 Oe from magnetization measurements on this sample. Importantly, the magnetization vs $H$ measurements of the $EuO_x$ overlayer are nearly identical for $H$ // [001] and [1$\bar{1}$0], implying that the observed MR anisotropy is inherent to the KTO 2DEG and not the magnetization of the $EuO_x$ overlayer. As the temperature is lowered towards $T_c$, the MR rapidly increases in magnitude, approaching ~ 130 % for $H$ // [001] , and ~ 47% for $H$ // [1$\bar{1}$0] for $H$ = 0.2 T at a temperature of ~ 1K. We also find that the highly anisotropic MR is independent of the current direction, from our measurements on Hall bars oriented along [001] and [1$\bar{1}$0] on the same sample, indicating that the anisotropy in the MR only depends on the orientation of the magnetic field with respect to the crystal lattice. We also note a weak anomaly in the temperature dependence of the MR near the Curie temperature of the $EuO_x$ overlayer (Fig. S3), further evidence for its coupling to electrons in the KTO 2DEG.

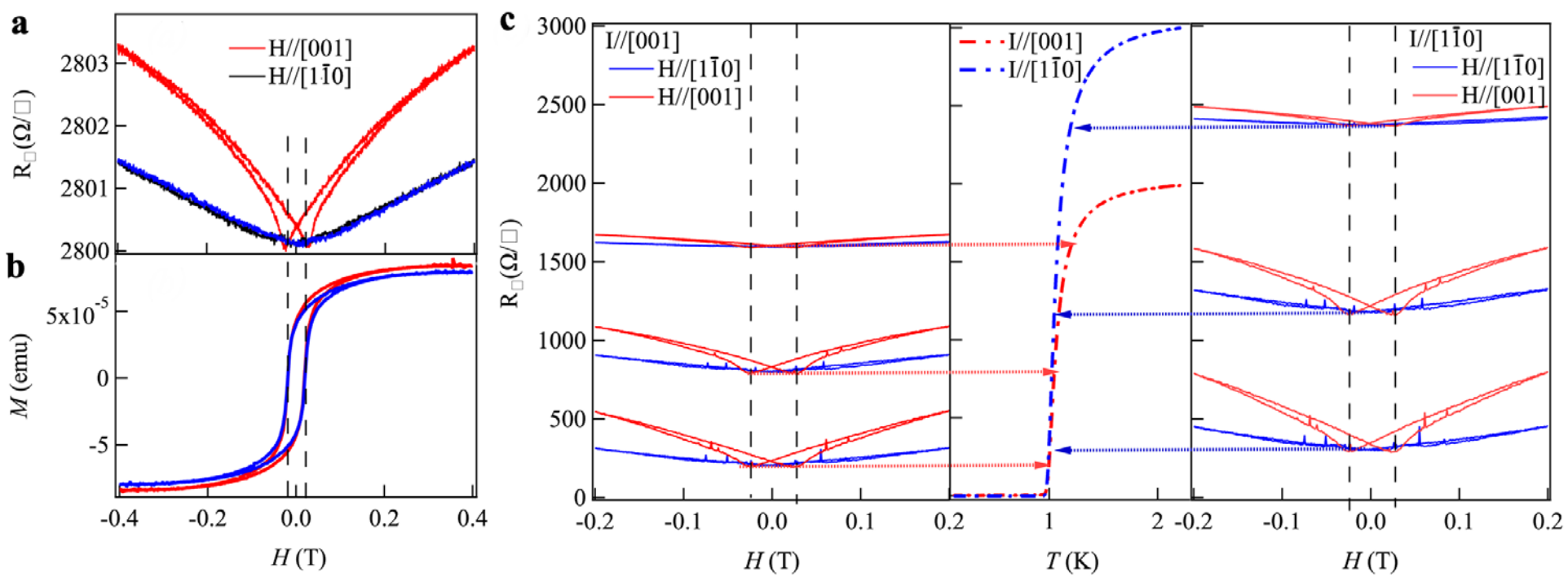

**Figure 2**. (a) Field dependence of the sheet resistance for EuO/KTO(110) at 2 K. The red solid line denotes a field along [001] and the blue solid line denotes a field along [1$\bar{1}$0]. (b) Field dependence of the magnetic moment for EuO/KTO(110) at 2 K, showing a nearly isotropic coercivity. The red solid line denotes a field along [001] and the blue solid line denotes a field along [1$\bar{1}$0]. (c) Field dependence of the sheet resistance for EuO/KTO (110) with current applied along different crystal axes. The left panel shows the field dependence of the sheet resistance when the current is applied along [001]. The right panel shows the field dependence of the sheet resistance when the current is applied along [1$\bar{1}$0]. The red solid line denotes a field along [001] and the blue solid line denotes a field along [1$\bar{1}$0]. The central panel shows the temperature dependence of the sheet resistance for I// [001] (red solid dashed line) and I// [1$\bar{1}$0] (blue solid dashed line). The arrows are used to indicate the different temperatures that the field dependence of the resistance is taken at.

We note that a hysteretic MR has been observed in 2DEGs formed at $LaAlO_3/SrTiO_3$, $GdTiO_3/SrTiO_3$ and $LaTiO_3/SrTiO_3$ interfaces when magnetism is induced [28-30] and recent work also shows hysteretic MR at KTO interfaces [31, 32]. However, the in-plane anisotropy in MR that we observe has not been reported earlier to the best of our knowledge. The anisotropy of the MR near the superconducting transition in Fig. 2 is consistent with a pair-breaking effect that arises from interfacial exchange coupling of ferromagnetic Eu spins in $EuO_x$ with the highly anisotropic in-plane uniaxial spin texture of KTO (110) 2DEGs. However, this low field MR for in-plane fields, including the hysteretic behavior, is no longer observed once we cool below $T_c$, as the experimentally measured in-plane $H_c$ increases far above the coercivity field of the $EuO_x$ overlayer. This also implies that the interfacial exchange coupling between the $EuO_x$ and the KTO (110) 2DEG is relatively weak, being effective only in the transition region near $T_c$ where the conductivity is enhanced by superconducting fluctuations, but not below $T_c$ where the superconductivity is more robust.

**Anisotropic in-plane upper critical field**

We also observe a striking anisotropy in the in-plane $H_c$ of $EuO_x$/KTO (110) 2DEGs as we cool below $T_c$. Figure 3(a)-(b) shows the temperature dependence of $R_\square$ when the field is applied along the [001] and [1$\bar{1}$0] directions. Figure 3(c) summarizes the $T_c$ dependence of $H_c$ for both directions. $T_c$ is taken as the inflection point in the temperature dependence of $R_\square$, normalized to its zero-field value. $H_c$ is normalized to the conventional Pauli paramagnetic field $\mu_0 H_P[T] = 1.86\, T_c\ [K]$ [33]. Note that the actual paramagnetic limiting field can be substantially higher both due to spin-orbit scattering and a reduced g-factor [34, 35]. At the lowest temperatures, in $EuO_x$/KTO (110) we find that $H_c$ // [1$\bar{1}$0] < $H_c$ // [001] where $\frac{H_c //\ [1\bar{1}0]}{H_c //\ [001]} \sim 0.8$ (Fig. 3, Fig. S9 (d)). However, closer to $T_c$, this ratio strongly *inverts* and $\frac{H_c //\ [1\bar{1}0]}{H_c //\ [001]} > 7$ at $T$ = 0.9 $T_c$. This switching of anisotropy in $H_c$ is observed in several $EuO_x$/KTO (110) samples, for a range of $T_c$ values (Figs. S5-7). The two-fold nature of the anisotropy in $H_c$ close to $T_c$ is evident in the angular dependence of the normalized $R_\square$ , $R_{ANI}(\varphi) = R_\square(\varphi)/R_\square(\varphi = 0)$, (here $\varphi = 0$ for $H$// [001]) in a 0.5 T in-plane magnetic field (Fig. 3(d)). A dumbbell-shaped $R_{ANI}(\varphi)$ is observed with minima for magnetic fields along [1$\bar{1}$0], implying $H_c$ // [1$\bar{1}$0] > $H_c$ // [001].

In addition, we also measured the dependence of $T_c$ on in-plane magnetic fields along [001] and [1$\bar{1}$0] for $AlO_x$/KTO (110) samples (Fig. S8). Figure 3(e) summarizes the $T_c$ dependence of $H_c$ for the [001] and [1$\bar{1}$0] directions for $AlO_x$/KTO (110) with $T_c$ = 1080 mK. In contrast to $EuO_x$/KTO (110), for $AlO_x$/KTO(110) the ratio of upper critical fields $\frac{H_c //\ [1\bar{1}0]}{H_c //\ [001]} \sim 0.98$ near $T_c$, and reduces monotonically to ~ 0.8 at the lowest temperatures, with no inversion. The same anisotropy is found in a lower $T_c$ sample (Fig. S9). We also measure $R_{ANI}(\varphi)$ in a 0.5 T in-plane (Figure 3(f)) for a fixed current direction to obtain the full angular dependence of MR at $T \lesssim T_c$ for $AlO_x$/KTO, and we find it to be nearly isotropic.

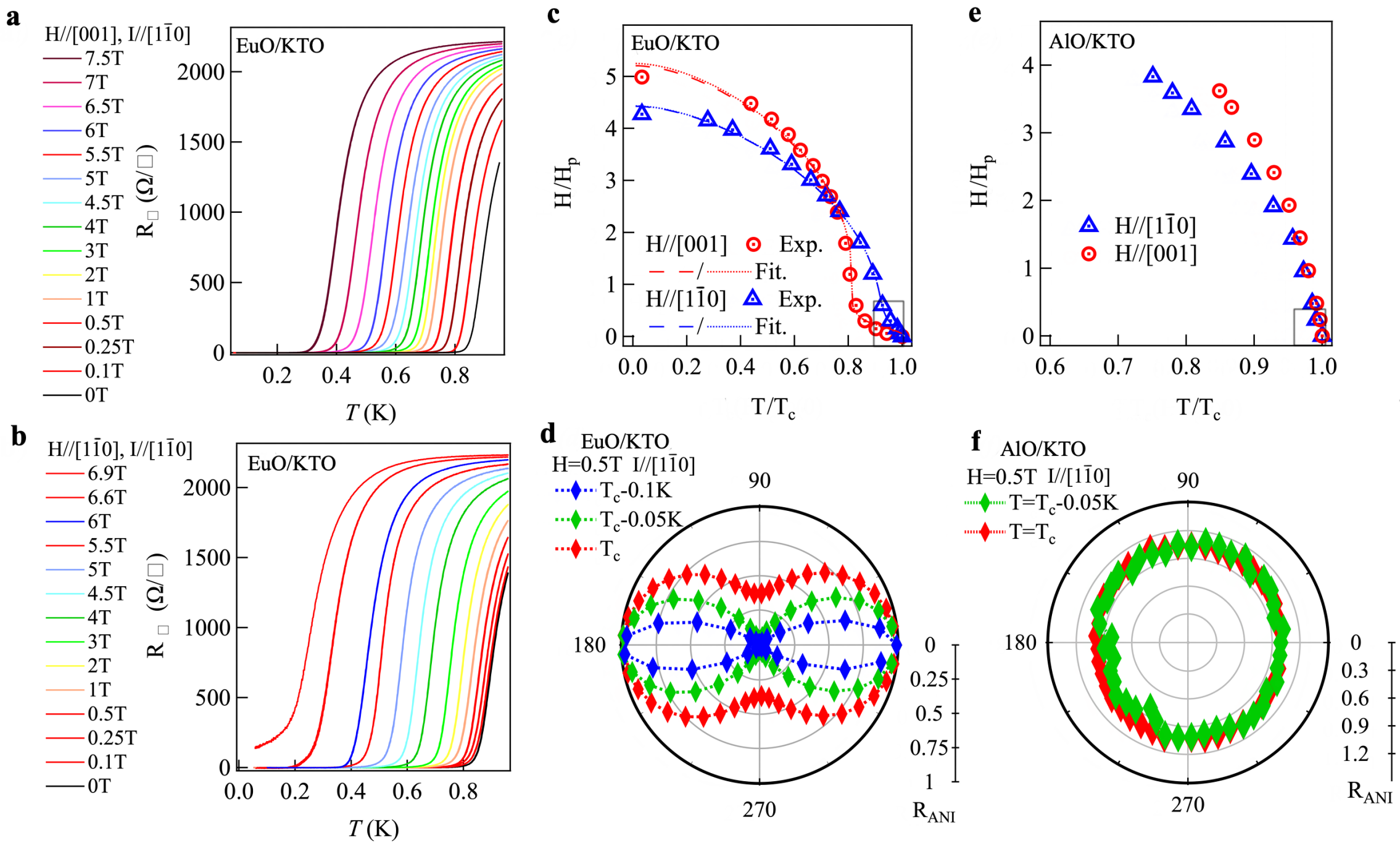

**Figure 3**. Temperature dependence of the sheet resistance of $EuO_x$/KTO for different in-plane fields when (a) the field is applied along [001] and (b) when the field is applied along [1$\bar{1}$0]. (c) Temperature dependence of the critical field of $EuO_x$/KTO for the [1$\bar{1}$0] and [001] field directions. $T_c$ is normalized to its zero-field value and $H_c$ is normalized to the Pauli paramagnetic field. The solid/dashed line denotes fits based on the KLB/WHH models described in the text. (d) The angular dependence of the normalized sheet resistance of $EuO_x$/KTO for an in-plane field of 0.5 T for three different temperatures. A lower $R_{ANI}$ would imply a higher value of $H_c$. (d) Temperature dependence of the critical field of $AlO_x$/KTO for the [1$\bar{1}$0] and [001] field directions. (e) The angular dependence with respect to the field direction of the normalized sheet resistance for $AlO_x$/KTO with a field of 0.5 T at two different temperatures. A higher $R_{ANI}$ would imply a lower value of $H_c$.

The $H_c$ vs $T_c$ phase diagrams for $EuO_x$/KTO (110) and $AlO_x$/KTO (110) are very different. Despite the highly anisotropic spin-texture anticipated for KTO (110) 2DEGs, the anisotropy observed in $AlO_x$/KTO (110) can be explained entirely by the anisotropy of the orbital-limiting field. A square-root like $H_c$ vs $T_c$ variation for $AlO_x$/KTO (110) is consistent with Ginzburg-Landau (GL) theory for a single band 2D superconductor with an anisotropic mass tensor [36], where the ratio of the orbital critical field along the two in-plane directions is approximately related to the inverse ratio of the effective masses. Our resistance measurements in the normal state are consistent with an effective mass $m_{[1-10]} > m_{[001]}$, resulting in $H_c$ // [001] > $H_c$ // [1$\bar{1}$0] as observed. Fits of the critical field using Werthamer-Helfand-Hohenberg (WHH) [37] theory for a lower $T_c$ sample with a wider dynamic range in temperature are consistent with a reduced *g*-factor (Fig.S16). While effective mass anisotropy is temperature independent, a reduced *g*-factor combined with anisotropic spin-orbit scattering can result in a modest variation of the anisotropy as a function of temperature (Fig.S16).

On the other hand, the unconventional $H_c$ vs $T_c$ data for $EuO_x$/KTO suggest contributions of both orbital and spin pair-breaking effects. We note that the hysteretic effects we observed at low fields of ~ 100 Oe for the data in Fig. 2 are absent at temperatures below $T_c$, and thus the observed effects are not due to ferromagnetic $EuO_x$. Rather, our data are reminiscent of Chevrel phase superconductors like $EuMo_6S_6$ where an internal field due to the presence of local moments has a profound impact on $H_c$ and can give rise to an inflection-like behavior near $T_c$ [38]. We suggest a scenario involving paramagnetic $Eu^{2+}$ impurity ions which diffuse into KTO, and interact via exchange interactions with the anisotropic spin texture of the KTO (110) 2DEG [17, 39, 40]. The external field rapidly polarizes the $Eu^{2+}$ spins at low temperatures, and they in turn interact with electrons at the Fermi surface via an exchange interaction. This gives rise to an extra pair-breaking term that acts to suppress $H_c$ for temperatures near $T_c$, that follows the polarization of the Eu moments. While an exchange coupling also exists between the ferromagnetic $Eu^{2+}$ in $EuO_x$ and conduction electrons in 2DEGs, the ferromagnetic moment rapidly saturate below 0.1 T; thus, it is insufficient to support the continued rapid change of critical field in $EuO_x$/KTO (110) for fields beyond 1 T along [001] near $T_c$. We note that the external magnetic fields interact via the Zeeman effect with the total moment ($L$+2$S$) of the electrons in the KTO (110) 2DEGs, which is strongly reduced due to the anti-aligned orbital and spin moments that nominally cancel. On the other hand, the exchange interaction between the $Eu^{2+}$ impurities and the conduction electron in the KTO (110) 2DEG involves only the electron spin ($S$), and is not affected by the reduced total moment.

To illustrate this, we generalize an approximate formula given by Klemm, Luther and Beasley (KLB) [41] by including an exchange field in their pair-breaking function $\alpha(h)$ that determines $H_c$:

$$ln(t) = \psi\left(\frac{1}{2}\right) - \psi\left(\frac{1}{2} + \frac{\alpha(h)}{2\pi k_B T}\right) \quad (1)$$

where $\psi$ is the digamma function, $t$ is the reduced temperature ($T/T_c$), and $h$ is the reduced field, which enters $\alpha(h)$ as $h^2$ as we are considering in-plane fields.

$$\alpha(h) = ch^2 + \frac{1}{2b}\left(\left(\frac{g}{2}\right)h + h_J\right)^2 \tag{2}$$

where $c$ is $\pi T_c$, $b^{-1}$ is $3\tau_{so}$ with $\tau_{so}$ the spin-orbit scattering time, and $h_J$ is the reduced exchange field between the paramagnetic $Eu^{2+}$ impurity ions ($J = 7/2$) and the tantalum conduction electrons which is described by a Brillouin function $h_J = h_{J0} B_J(Jg'\mu_B H/k_B T)$ [30] ($g'$= 2). Here, $h_{J0}$ is the magnitude of the exchange field depending on the microscopic exchange integrals and the concentration of Eu-ions. $h_{J0}$ also depends on the angle between the in-plane field $h$ and the [001] axis due to the anisotropy of the spin texture of the Ta-5$d$ states. Since the spin texture in Fig. 1(b) has a uniaxial in-plane anisotropy, $h_J$ is larger for fields along [001] compared to fields along [1$\bar{1}$0]. Thus, when the external field is applied along the [001] direction, the pair-breaking function $\alpha(h)$ will rapidly increase at small fields following the H/T dependence of the Brillouin function. Due to this 'boost' to $\alpha(h)$ provided by the exchange field, the required external field for pair-breaking is greatly reduced, and $H_c$// [001] is much smaller for $T$ near $T_c$. Once the moment is saturated, the relative contribution of the 'exchange' term is diminished as the field increases further. Therefore, with the decrease of temperature and correspondingly the increase in critical field, $H_c$ *vs* $T_c$ is gradually dominated by pair-breaking due to orbital effects. For this reason, both $EuO_x$/KTO and $AlO_x$/KTO show similar behavior at low $T$. We extract $\alpha(h)$ from our experimental $H_c$ and fit it using Eq. (2) assuming $g = 0.5$ (Fig. S17) and find that $\frac{h_J\ [001]}{h_J\ [1\bar{1}0]} \sim 1.33$. An alternative fit to the data can be done by using an in-plane formula for $H_c$ which is based on WHH supplemented by the exchange field [42] (Fig. S17). As with the KLB fit, the exchange field has a negative sign with respect to applied field in the WHH fit, and a similar ratio of $\frac{h_J\ [001]}{h_J\ [1\bar{1}0]} \sim 1.29$ is obtained. While a non-zero $g$-factor implies a Zeeman contribution to pair-breaking from the applied field, the exchange field remains the dominant pair-breaking factor near $T_c$.

For context, the crossover in the anisotropy in $H_c$ has been observed [43-45] in a quasi-1D superconductor $K_2Cr_3As_3$, which arises from an anisotropic spin susceptibility along and perpendicular to its 1D chains, though these data do not show the pronounced inflection in $H_c(T_c)$ we see for fields along [001] in $EuO_x$/KTO (110). Notably, measurements of the Knight shift in $K_2Cr_3As_3$ in the superconducting state suggest a triplet component [46]. Two-fold modulations of the in-plane $H_c$ have also been observed in several 2D superconductors, and have been attributed to emergent nematicity, [47] and mixing of *s*-wave with *d*-wave or *p*-wave order parameters [11, 48]. While a Lorentz force resulting from magnetic fields could lead to a two-fold symmetry of the in-plane $H_c$ if vortices move freely, the observed results in $EuO_x$/KTO are independent of the current direction (Fig. S10) which would rule out this mechanism for an anisotropic Fermi surface. More recently, two-fold $H_c$ modulations have been ascribed to anisotropic spin susceptibilities that result from 3D spin textures of the Fermi surface due to broken mirror symmetries, and the interplay of parity with spin-orbit coupling near band-inversion points in reciprocal space [49-51]. However, corroborating experimental evidence for these spin textures is not well established.

**Diffusion of Eu ions across the KTO (110) interface**

Scanning transmission electron microscopy (STEM) measurements yield information on the distribution of Eu ions in the vicinity of the $EuO_x$/KTO (110) interface, where cation interdiffusion and presence of O vacancies can lead to the formation of a 2D electron gas [40]. Figure 4 (a) shows a cross-sectional high-angle annular dark field (HAADF) image along the KTO [110] zone-axis orientation, where the Eu ions that have diffused into KTO are evident from the enhanced Z-contrast on the K sites (Fig. 4(b), Fig. S11), up to several atomic layers into the KTO. To obtain further insight regarding diffusion of Eu, we also analyze the interfacial composition using energy-dispersive x-ray spectroscopy (EDS) (Fig.4(c)). The concentration of Eu decreases as it goes from $EuO_x$ into KTO, and the presence of Eu persists beyond 5 nm into KTO (Fig. 4(d)). Electron energy loss spectroscopy (EELS) of Eu also indicates diffusion of Eu into the KTO (Fig. 4(e)). In particular, the Eu fine-structure measured using EELS reveals new satellite peaks that emerge when scanning from $EuO_x$ into the KTO (Fig. 4(e)). The intensity of the satellite peak at 1133 eV corresponding to $Eu^{3+}$ rises near the interface before decaying in the KTO within 6.8 ± 1.0 nm, while the spectral signatures for $Eu^{2+}$ peaked at 1135.2 eV decay smoothly going from $EuO_x$ into the KTO within about 3.7 ± 1.1 nm. This is within the effective thickness of the superconducting electron gas at the $EuO_x$/KTO interface of ~ 7 nm (Fig.S12). Thus, our findings are consistent with a mixture of $Eu^{2+}$ and $Eu^{3+}$ ions within KTO near the interface (Fig. 4f). The presence of $Eu^{2+}$, which has a magnetic moment of $7\mu_B$, is consistent with the field dependence of the pair-breaking parameter that we observe at temperatures just below $T_c$. On the other hand, $Eu^{3+}$ has zero net moment, and will not respond to external fields. Thus, we hypothesize that when an external field is applied, paramagnetic $Eu^{2+}$ ions within the KTO align with the field and interact with the electrons in the KTO (110) 2DEG through exchange coupling.

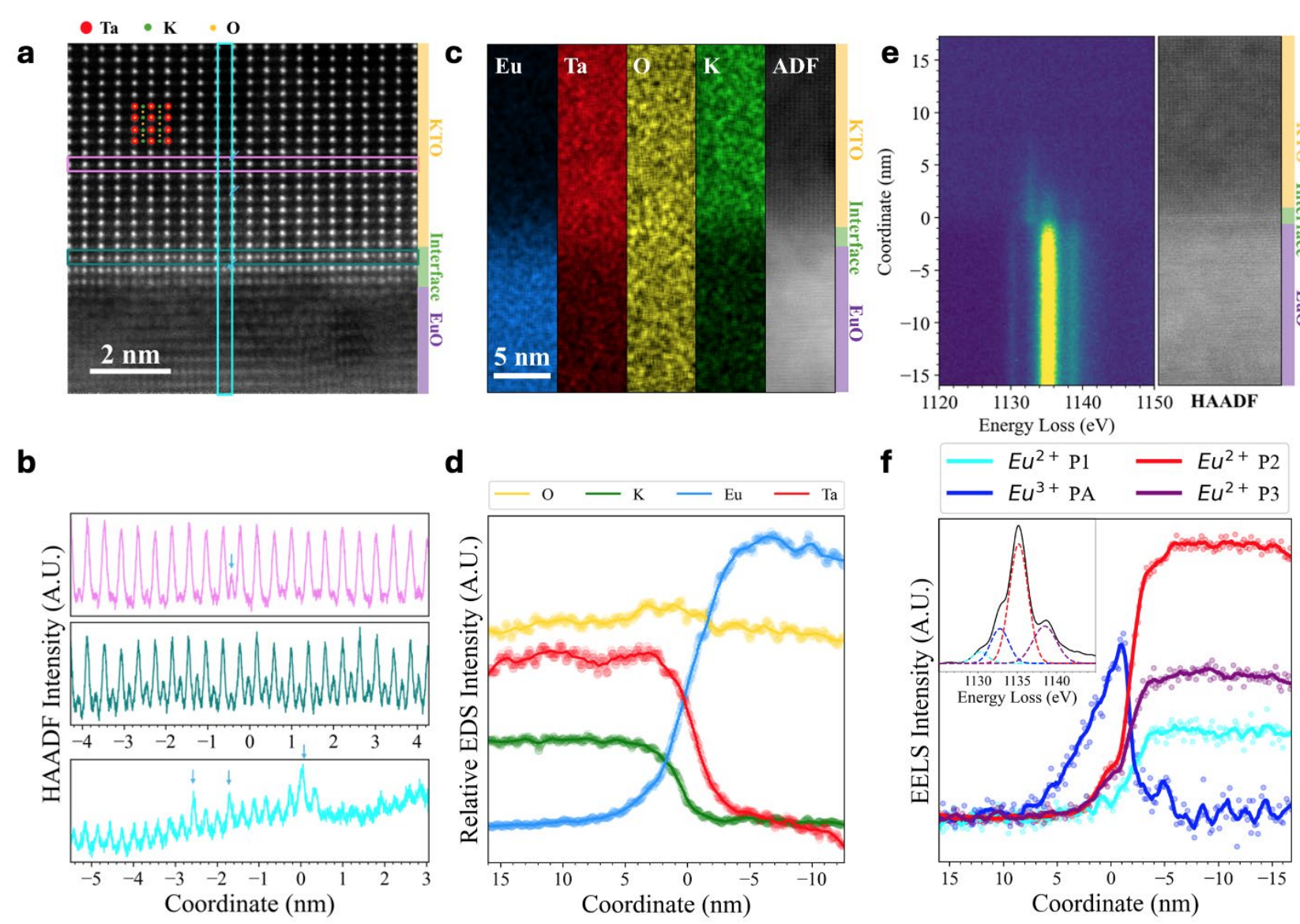

**Figure 4**. a) Atomic resolution HAADF image of a EuO/KTO [110] interface. b) Intensity line profile of the three colored lines marked in a), the higher intensity column of atoms marks the replacement of K by Eu. c) Larger field of view EDS mapping and HAADF image across the interface. d) EDS intensity profile of K, Ta, O and Eu with respect to the vertical direction in c) from KTO to EuO across the interface. e) 2D EELS spectra (left panel) correlated to the HAADF image (right panel) at the EuO/KTO [110] interface. f) Spatial profiles for the $Eu^{2+}$ (1130.3eV, 1135.2eV and 1138.5eV) and $Eu^{3+}$ (1132.8eV) peaks in the EELS spectra measured across the EuO/KTO [110] interface. A selected spectrum and its decomposed peaks are shown in the inset. The zero position in (b), (d) and (f) is the EuO/KTO [110] interface.

## Anisotropic weak antilocalization for in-plane magnetic fields

To further corroborate the uniaxial in-plane spin texture, we measure magnetoconductance at $T > T_c$ for in-plane fields up to 5T, where we find signatures of an anisotropic in-plane weak antilocalization (WAL) in $EuO_x$/KTO (110) samples. Quantum interference of conduction electrons in the presence of SOC results in a positive magnetoresistance known as WAL. WAL for an out-of-plane field has been observed for 2DEGs in both KTO (111) and KTO (110) [34, 52-54] . However, in 2DEGs formed at the $NdTiO_3/SrTiO_3$ interface, it was shown that exchange coupling due to local magnetic moments can enhance the dephasing of electrons under *in-plane* fields [55], and that WAL effects in this geometry can probe the in-plane spin susceptibility. Thus, we first measured the magnetoconductance in $AlO_x$/KTO (110) at different temperatures for in-plane fields along [1$\bar{1}$0] and [001] (Fig. 5(a)-(c)). A broad quadratic field dependence of the magnetoconductanceis observed along both directions, with a decrease in the magnetoconductance as the temperature increases due to the loss of phase coherence in the field. A cusp-like behavior is only observed for fields along the out-of-plane direction (Fig. S13). This mirrors what occurs for $H_c$, where the orbital field enters linear in $H$ for perpendicular fields but quadratic in $H$ for parallel fields, given that the Cooperon that is associated with WAL has similar properties to the Cooper pair propagator for superconductivity.

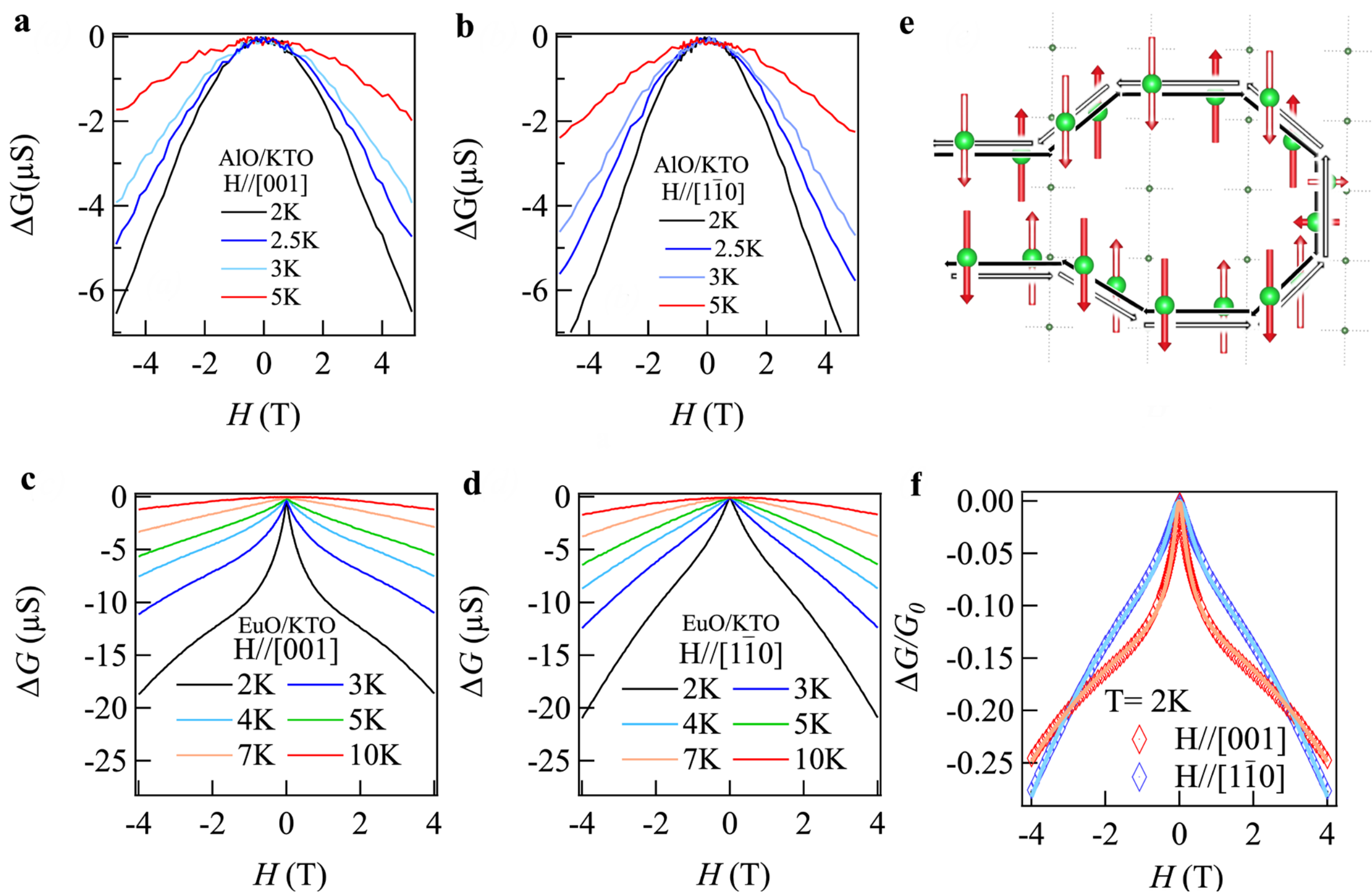

**Figure 5** Field dependence of the magnetoconductance at different temperatures for $AlO_x$/KTO for an in-plane magnetic field applied (a) along the [001] direction and (b) along the [1$\bar{1}$0] direction. (c) and (d) are the same as (a) and (b) but for $EuO_x$/KTO. (e) Schematic of the destructive interference for an electron scattering in time-reversed closed loop paths in KTO(110) with spins (shown by solid and hollow red arrows) locked along the [001] axis. (f) Weak anti-localization fits to the magnetoconductance of $EuO_x$/KTO at 2 K for both in-plane field directions as described in the text.

Figure 5 (c)-(d) shows the magnetoconductance measured on $EuO_x$/KTO (110) at different temperatures for the two in-plane field directions. In strong contrast to $AlO_x$/KTO (110), a sharp cusp in the field dependence of the magnetoconductance is observed when the field is applied along [001]. As the temperature increases from 2 K, the cusp feature is attenuated and the magnetoconductance decreases sharply as quantum corrections are lost. For fields applied along [1$\bar{1}$0], a noticeably broader cusp-like feature is observed at 2 K, but with a similar temperature dependence as for [001]. Since the orbital corrections to the magnetoconductance should be quadratic and small for in-plane fields in a quasi-2D limit, we look to the Zeeman contribution for the dephasing of WAL in the magnetoconductance [55],

$$\Delta G(H_{//}) = -\frac{G_0}{2} ln\left(1 + \frac{\Delta_\Phi(H_{//})}{B_\Phi}\right) \tag{3}$$

where $\Delta_\Phi(H_{//}) = \frac{[2\mu_B(H_{//})]^2}{(4eD)^2 B_{so}}$. Here, $D$ is the diffusion constant, $B_{so}$ is the field scale associated with spin-orbit scattering, $B_\Phi$ is the orbital dephasing field scale, and $G_0$ is the quantum of conductance. For an in-plane field *H,* this gives rise to a quadratic field dependence. However, as with $H_c$, for $EuO_x$/KTO, we expect that the Eu impurity moments in the KTO 2DEGs will induce an additional exchange term that will sharply enhance the effective field that enters into the Zeeman term [56] $H_{//}=\frac{g}{2}H+H_J$ . The first term in $H_{//}$ is small since $g$ is strongly reduced due to quenching of the total moment, and the field-induced dephasing is dominated by the exchange interaction between Eu moments and the 'in-plane' uniaxial spin polarized electrons (Fig. 5 (e)). At low temperatures, the Eu moments align rapidly with $H$ leading to the pronounced 'boost' in the exchange interaction at low fields with electrons in the KTO 2DEGs, and a quasi-linear behavior is seen in the magnetoconductance. At higher fields, the impurity moments saturate, and we anticipate crossing over to a quadratic behavior as expected from orbital contributions. We thus fit our data using Eq. 3 with the same $g$ and the same exchange field as in our $H_c$ analysis (See SI Table 2). These fits are excellent, indicating that our $H_c$ and magnetoconductance analysis are consistent with one another.

**Conclusions**

In summary, we find evidence for an in-plane uniaxial spin texture in KTO (110) 2DEGs. The texture is hidden in $AlO_x$/KTO, due to a strongly reduced *g*-factor of the Ta conduction electrons, but for $EuO_x$/KTO the presence of an exchange field from the Eu ions 'lights up' the Ta-*5d* spin texture and allows us to see its consequences for both the critical field in the superconducting state as well the low-temperature normal state magnetoconductance. The anisotropic response in the superconducting state is consistent with the uniaxial 'half-Rashba' ($\boldsymbol{k}_y \hat{\sigma}_x$) spin texture that occurs in the $d_{yz}/d_{xz}$ bands. Our results also imply that carriers in $d_{xy}$ bands, which lack such a uniaxial spin texture, do not contribute. These carriers also do not participate in the weak antilocalization like response to in-plane fields. In addition to revealing the spin texture at KTO (110) interfaces, the magnetic proximity effects offer a path to control superconductivity, which can be further explored by using different magnetic overlayers in future studies of these heterostructures. The unique in-plane half-Rashba like nature of the spin texture intrinsic to all KTO (110) 2DEGs may have consequences both for applications in spintronics [57], and the nature of superconductivity in these 2DEGs. In particular, the triplet component of the Cooper pairs that can be admixed due to inversion symmetry breaking should exhibit the same in-plane uniaxial anisotropy as we find here.

**Acknowledgements**

We thank Ivar Martin for discussions. All research presented here is supported by the Materials Science and Engineering Division, Office of Basic Energy Sciences, U.S. Department of Energy. Work performed at the Center for Nanoscale Materials, a U.S. Department of Energy Office of Science User Facility, was supported by the U.S. DOE, Office of Basic Energy Sciences, under Contract No. DE-AC02-06CH11357. C.L. acknowledges partial financial support from the College of Arts and Sciences, University at Buffalo, SUNY. The National High Magnetic Field Laboratory is supported by the National Science Foundation through NSF/DMR-1644779, NSF/DMR-2128556, and the State of Florida.

**Contributions**

Synthesis of samples was carried out by J.Y. and C.L. with assistance from A.B. and J.P. All low-temperature magneto-transport and magnetization measurements were carried out by J.Y. and C.L. with assistance from U.W., A.S., J.S.J., X.J. and D.J. The theory analysis presented here was carried out by M.R.N. along with J.Y. and A.B. The paper was written by J.Y., M.R.N. and A.B. and all authors contributed to discussions regarding the paper.

## Statement regarding error bars

In Figures.1,2,3 and 5, the uncertainty/error in the plotted data points are smaller than the symbols used to represent the data. For Figure 4, the error bars for the diffusion lengths of the $Eu^{2+}$ and $Eu^{3+}$ in KTO are estimated from the scatter in the intensity of the spectral signatures with position.

## Data availability

All data for figures in the manuscript will be made available via the Harvard Dataverse.

# Supplemental Information

# Uniaxial spin texture in a superconducting electron gas revealed by exchange interactions

Junyi Yang[1*], Changjiang Liu[1,2*], Xianjing Zhou[3], Hanyu Hou[4], Kaijun Yin[4], Jianguo Wen[3], John Pearson[1,3], Alexey Suslov[5], Dafei Jin[3,6], Jidong S. Jiang[1], Ulrich Welp[1], Jian-Min Zuo[4], Michael R. Norman[1], Anand Bhattacharya[1]

1. Materials Science Division, Argonne National Laboratory, Lemont, USA
2. Department of Physics, University at Buffalo, SUNY, Buffalo, USA
3. Center for Nanoscale Materials, Argonne National Laboratory, Lemont, USA
4. Department of Material Science and Engineering, University of Illinois at Urbana-Champaign, Urbana, USA
5. National High Field Magnetic Laboratory, Tallahassee, USA
6. University of Notre Dame, Notre Dame, USA

## Methods

*Sample Fabrication*

$AlO_x$/KTO samples were prepared by depositing pure Al on a single crystal KTO (110) substrate in a temperature range from 500 C to 600 C under $10^{-9}$ mbar pressure in a molecular beam epitaxy chamber. The $EuO_x$/KTO sample was prepared by depositing pure Eu on a single crystal KTO (110) substrate at a temperature of 500 C under $10^{-8}$ mbar oxygen pressure in a molecular beam epitaxy chamber[1, 2]. A 10 nm Ge capping layer was deposited on all samples as a protecting layer. Hall bar patterns were prepared using by means of a maskless laser photo-lithography system (MLA 150, Heidelberg; photoresist AR-P 5350, Allresist EN). Pattern transfer from the photoresist onto the samples was realized using liq-$N_2$ cooled Ar-based ion-beam etching.

*Transport measurements*

The low-temperature transport measurements down to 50 mK were carried out in a Bluefors dilution fridge equipped with a 9-1-1 vector magnet. The out-of-plane magnetic field component for a given in-plane magnetic field was zeroed out by searching for minima in the magnetoresistance while sweeping the out-of-plane field about zero.

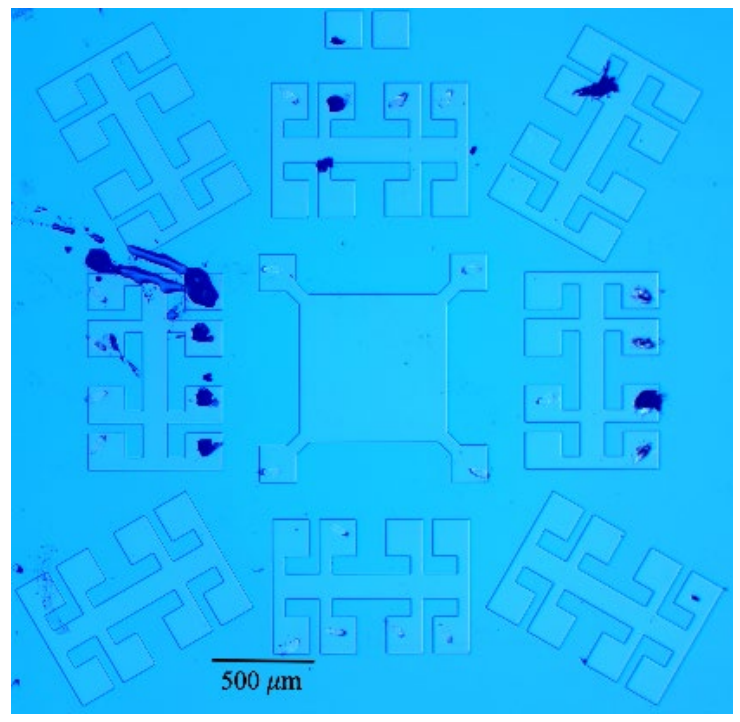

**Figure S1**. Devices fabricated on KTO (110) under an optical microscope. The black dots are wire bonder residuals after measurements.

The 23 mK field dependence of the resistance measurement in parallel fields was carried out in an 18 T cryogenic station at the National High Magnetic Field Laboratory, with the sample mounted on a rotator. The zeroing out of the out-of-plane component was achieved by searching for minima in the angular dependence of the resistance while rotating the sample. The magnetoconductance measurements at higher temperature were carried out through a rotator in a 14 T Physical Properties Measurement System from Quantum Design.

## S1: Patterned 2DEG KTO devices

Hall bar devices were fabricated on a KTO (110) sample along different crystal axes. A square device was patterned in the center for a Van der Pauw geometry measurement (Figure S1).

## S2: XRD measurement for in-plane orientations

The patterned thin film sample was mounted on the dilution fridge puck and the puck was mounted on the sample stage of a X'Pert X-ray diffractometer (Fig. S2 (a)). The $\theta - 2\theta$ measurement was first carried out along the out-of-plane direction where a clear [110] KTO Bragg peak was observed (Fig. S2 (b)). Due to the absence of a monochromator on the diffractometer, Bragg diffractions due to both Cu K-$\alpha$ and Cu K-$\beta$ were present. To resolve the in-plane orientation of the sample, a survey of different diffraction peaks was conducted. The Bragg peak [221], which is equivalent to [001]+2[110], was observed at Phi = 0° (Fig. S2 (c)). The Phi dependence of the Bragg peak [221] shows that it is only observed for Phi = 0° and 180° (Fig. S2 (d)). The peak intensities at Phi = 0 and 180° differ because of slight off-axis alignment. On the other hand, the Bragg peak [310], which is equivalent to [1-10]+2[110], is observed at Phi = 90° (Fig. S2 (e)). The Phi dependence of the Bragg peak [310] shows that it is only observed at Phi = 90° and 270° (Fig. S2 (f)). To avoid confusion of orientation in transport measurements, the sample was kept on the same puck and directly mounted on to the dilution fridge.

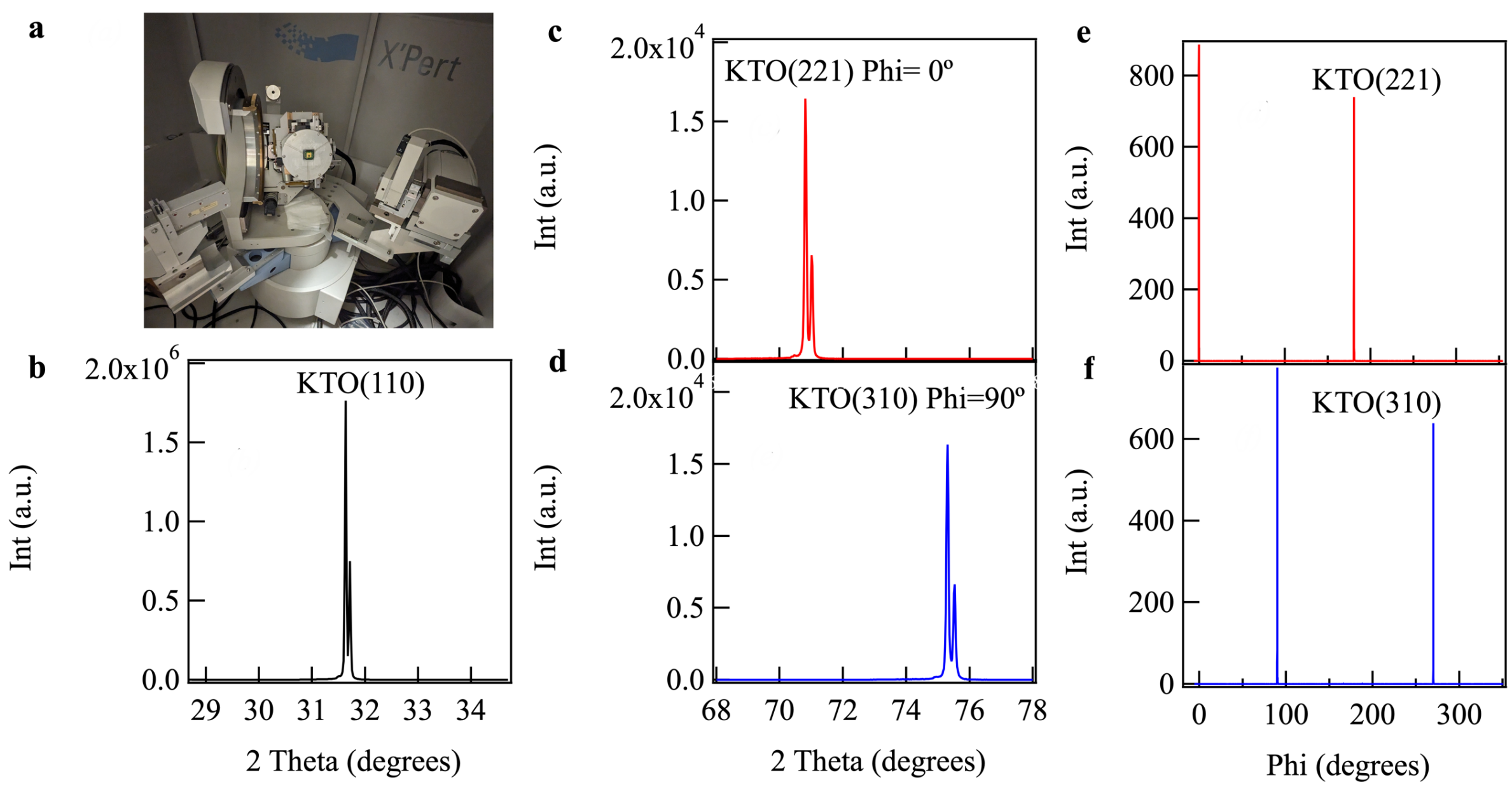

**Figure S2**. (a) Schematic of the XRD measurement setup. (b) $\theta - 2\theta$ scan around KTO (110). (c) $\theta - 2\theta$ scan around KTO (221). (d) Phi dependence of the intensity of KTO (221). (e) $\theta - 2\theta$ scan around KTO (310). (f) Phi dependence of the intensity of KTO (310).

## S3: Magnetoresistance for $EuO_x$/KTO around the EuO magnetic transition

The field dependence of the magnetoresistance for $EuO_x$/KTO for temperatures in the range of 50-80 K is shown in Fig. S3 (a) with H // [001] (in plane). The field dependence of the magnetoresistance at low fields is quadratic for a temperature (80 K) above the Curie temperature of EuO, and it becomes quasi-linear for temperatures below this transition. The temperature dependence of the magnetoresistance is shown in Fig. S3 (b). The magnetoresistance is defined as [R(-3T)-R(0T)]/R(0T) and it is negative for all measured temperatures. While a change in slope is observed close to the Curie temperature of EuO, the overall magnetoresistance is less than 0.1%. While the small negative magnetoresistance is consistent with a previous report on a EuO/KTO (001) interfacial electron gas[3], we also note that we do not observe any evidence for an insulator-metal transition in our $EuO_x$ overlayer as has been observed in EuO thin films[4].

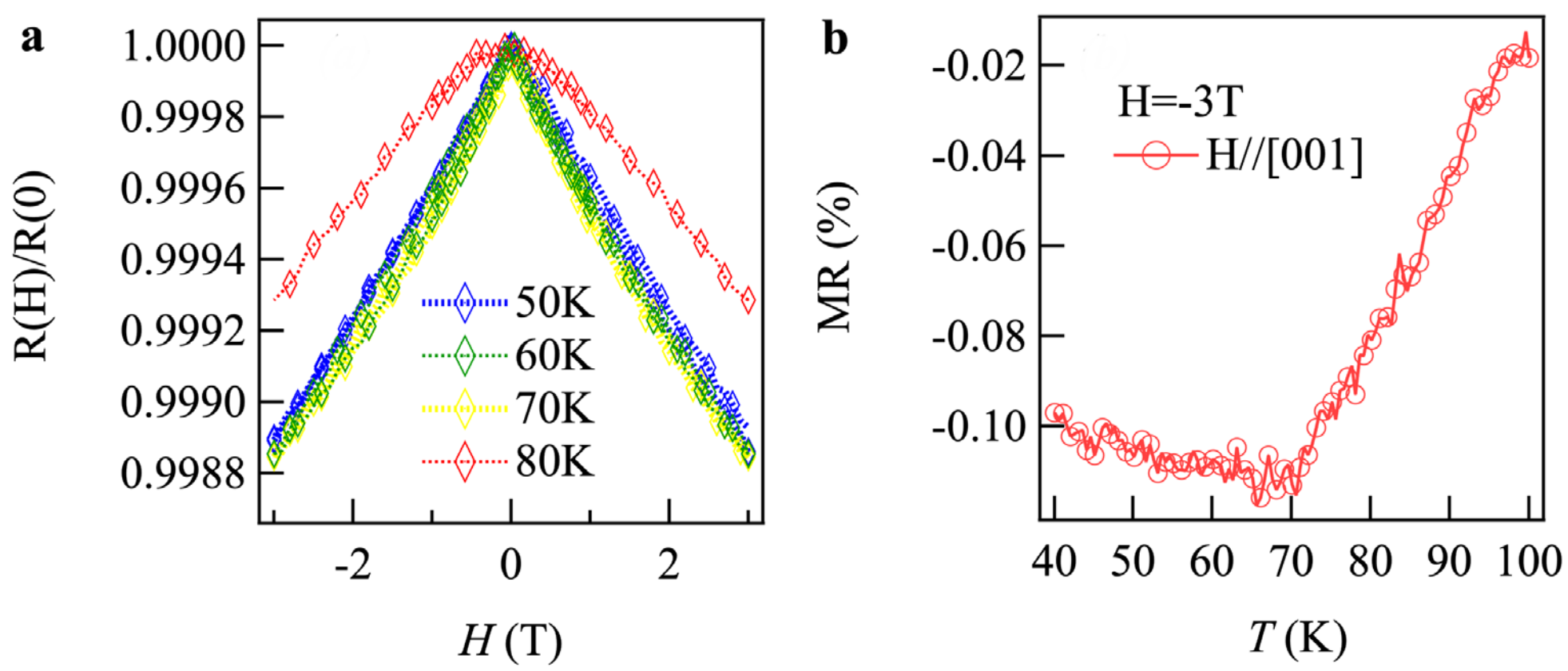

**Figure S3** (a) Field dependence of the resistance of $EuO_x$/KTO at various temperatures for H//[001]. (b) Temperature dependence of the magnetoresistance of $EuO_x$/KTO for a -3T field. Note the anomaly at the Curie temperature of EuO.

## S4: Field dependence of the resistance of $EuO_x$/KTO

The field dependence of the resistance of $EuO_x$/KTO at 23 mK is shown in Fig. S4. The field is applied along [001]/[1-10] respectively.

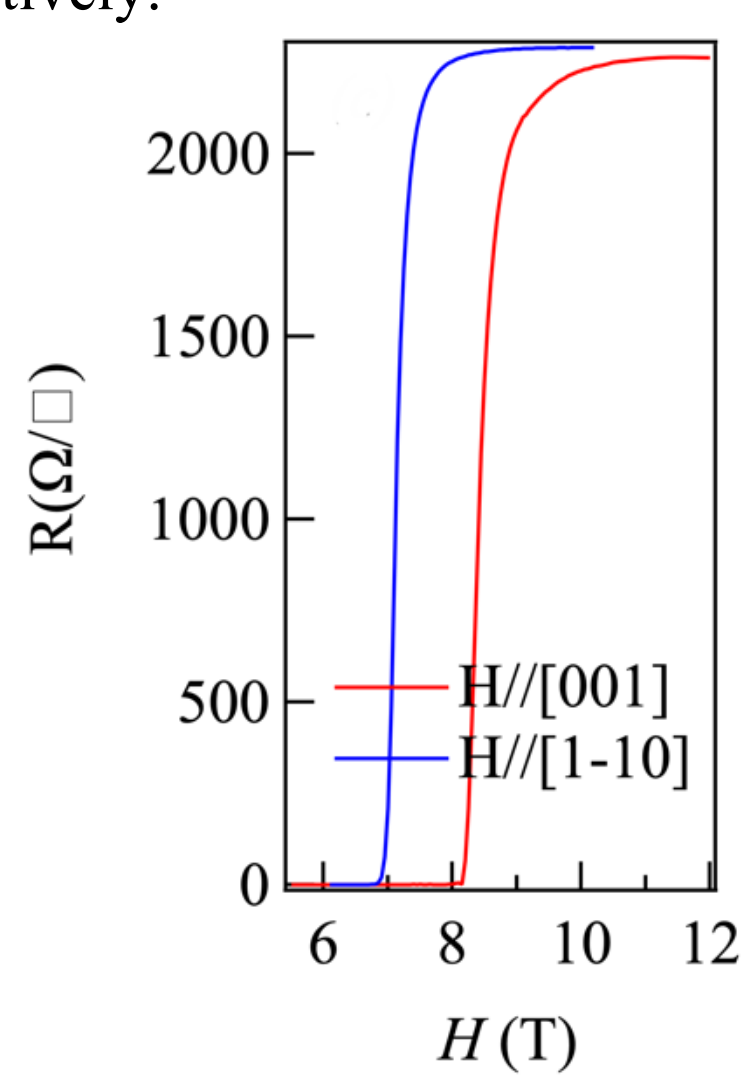

**Figure S4.** Field dependence of the resistance of $EuO_x$/KTO(110) with a field along [001] (red) and [1-10] (blue).

## S5&S6: In-plane anisotropy for $EuO_x$/KTO samples with $T_c$ = 1050 mK and $T_c$ = 890 mK

The temperature dependence of the resistance under in-plane fields along [001] and [1-10] is shown in Fig. S5 (a) and (b). The transition temperature is about 890 mK. The critical field vs temperature is extracted from Fig. S5 (a) and (b) and presented in Fig. S5 (c). The critical field is larger along [001] at low temperatures, while the critical field along [1-10] is larger at temperatures close to $T_c$. A similar measurement is carried out for sample with $T_c$ = 1050 mK.

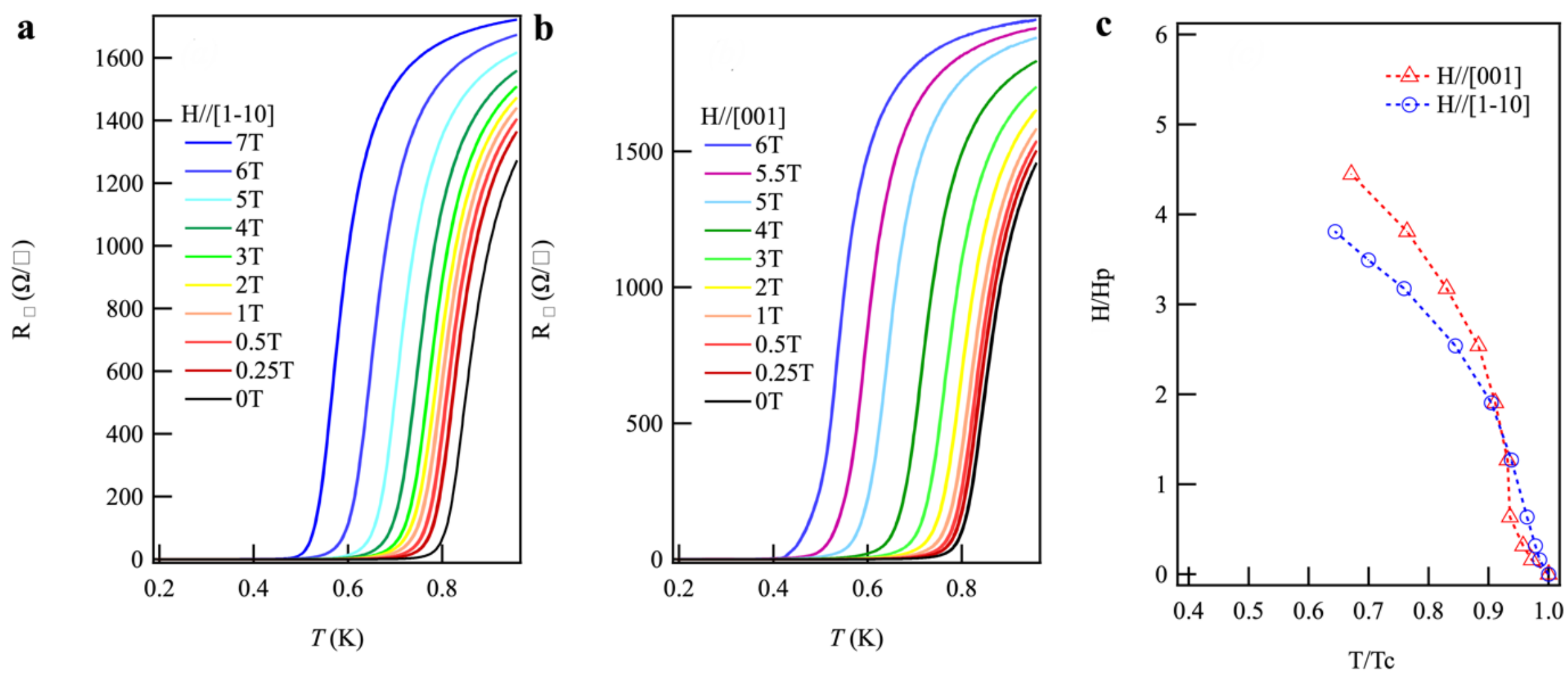

**Figure S5**. $EuO_x$/KTO sample with $T_c$ = 1050 mK (a) Temperature dependence of the resistance for different H//[001]. (b) Temperature dependence of the resistance for different H//[1-10]. (c) Critical field vs critical temperature for the two field directions.

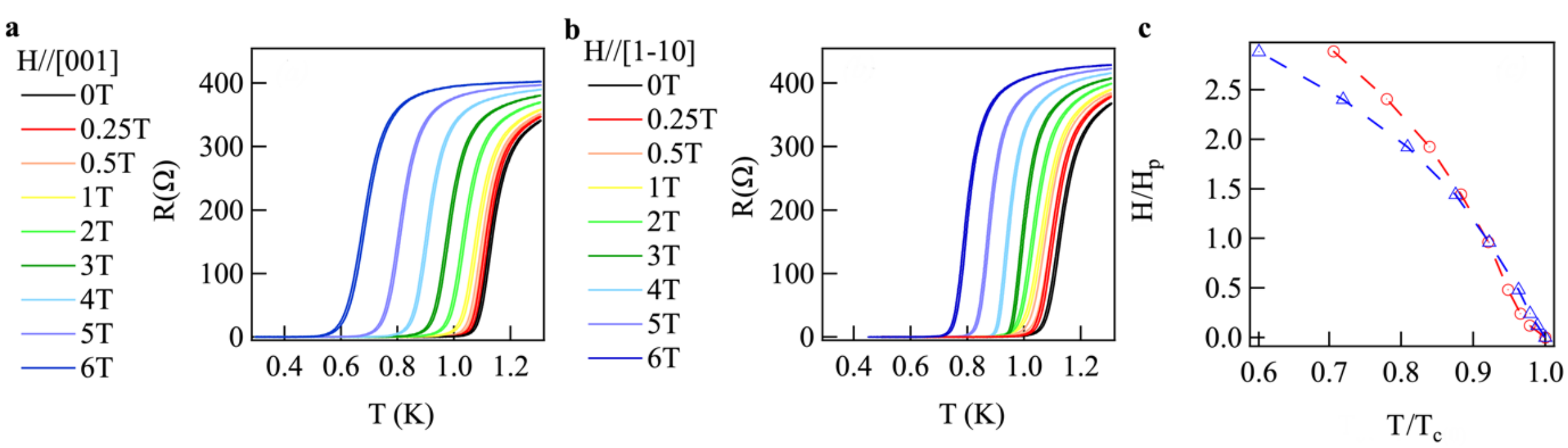

**Figure S6**. $EuO_x$/KTO sample with $T_c$ = 890 mK Temperature dependence of the resistance for different H//[001]. (b) Temperature dependence of the resistance f different H//[1-10]. (c) Critical field vs critical temperature for the two field directions.

## S7: In-plane anisotropy for a $EuO_x$/KTO sample with $T_c$ = 275 mK

The temperature dependence of the resistance under in-plane fields along [001] and [1-10] is shown in Fig. S7 (a) and (b). The transition temperature is about 275 mK. The critical field vs critical temperature is extracted from Fig. S7 (a) and (b) and presented in Fig. S7 (c). The critical field is larger along [001] at low temperatures, while the critical field along [1-10] is larger at temperatures close to $T_c$. The angular dependence of the resistivity with respect to the in-plane field direction at a temperature close to $T_c$ is presented in Fig. S7 (d). The resistance for a field along [1-10] is smaller than that along [001], which is consistent with the higher $T_c$ $EuO_x$/KTO sample presented in the main text. While the anisotropic behavior remains for this lower $T_c$ $EuO_x$/KTO sample, its anisotropy is smaller.

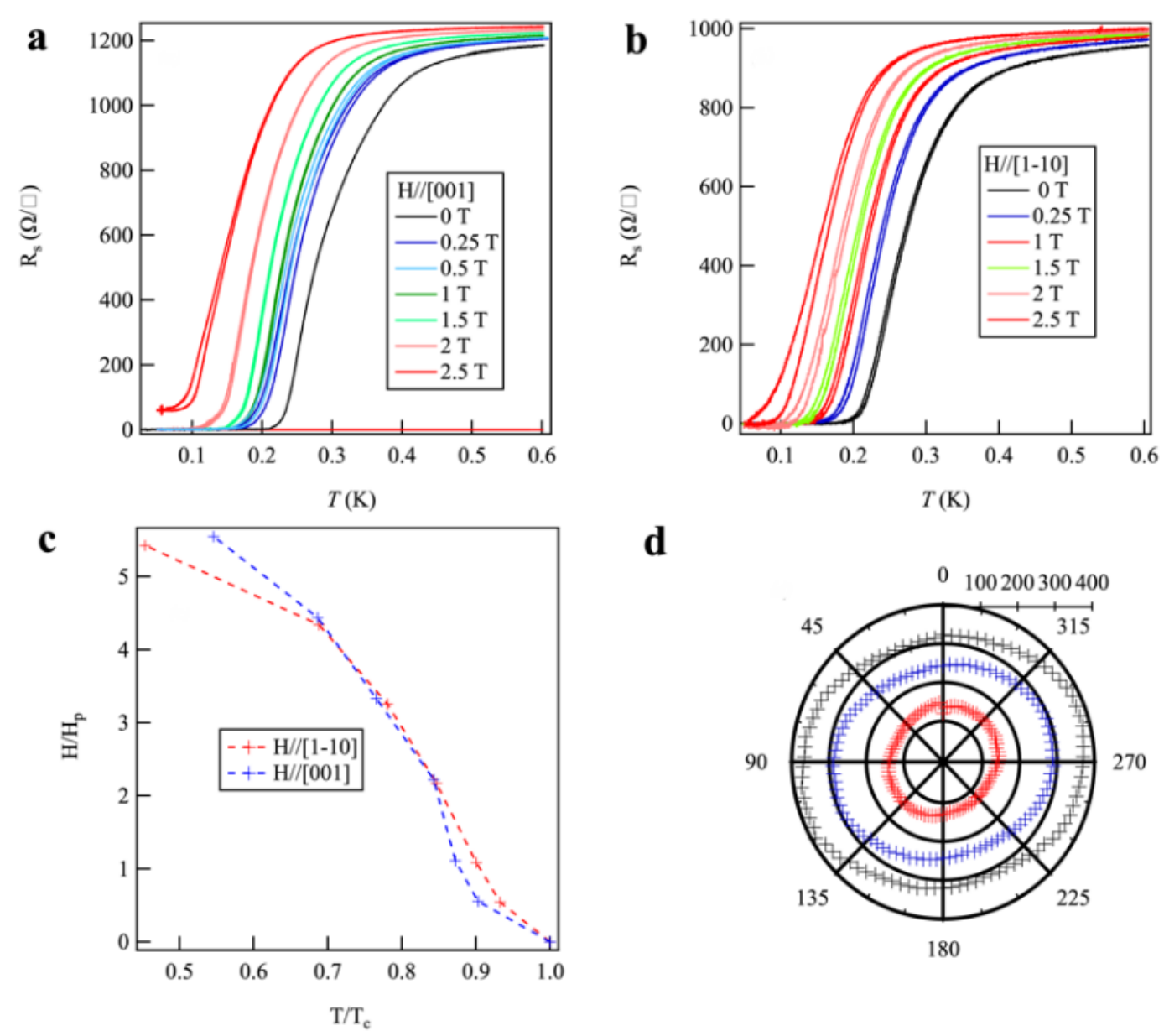

**Figure S7.** (a) Temperature dependence of the resistance for different H//[1-10] for $EuO_x$/KTO with $T_c$~ 275 mK. (b) Temperature dependence of the resistance for different H//[001]. (c) Critical field vs critical temperature for the two field directions. (d) Angular dependence of the resistance versus the in-plane field direction at 255 mK. The black curve denotes the field of 0.5T, the blue curve denotes the field of 0.25T and the red curve denotes the field of 0.1T.

## S8: In-plane anisotropy for $AlO_x$/KTO sample with $T_c$ = 1080 mK

The temperature dependence of the resistance under in-plane fields along [001] and [1-10] is shown in Fig. S8 (a) and (b).

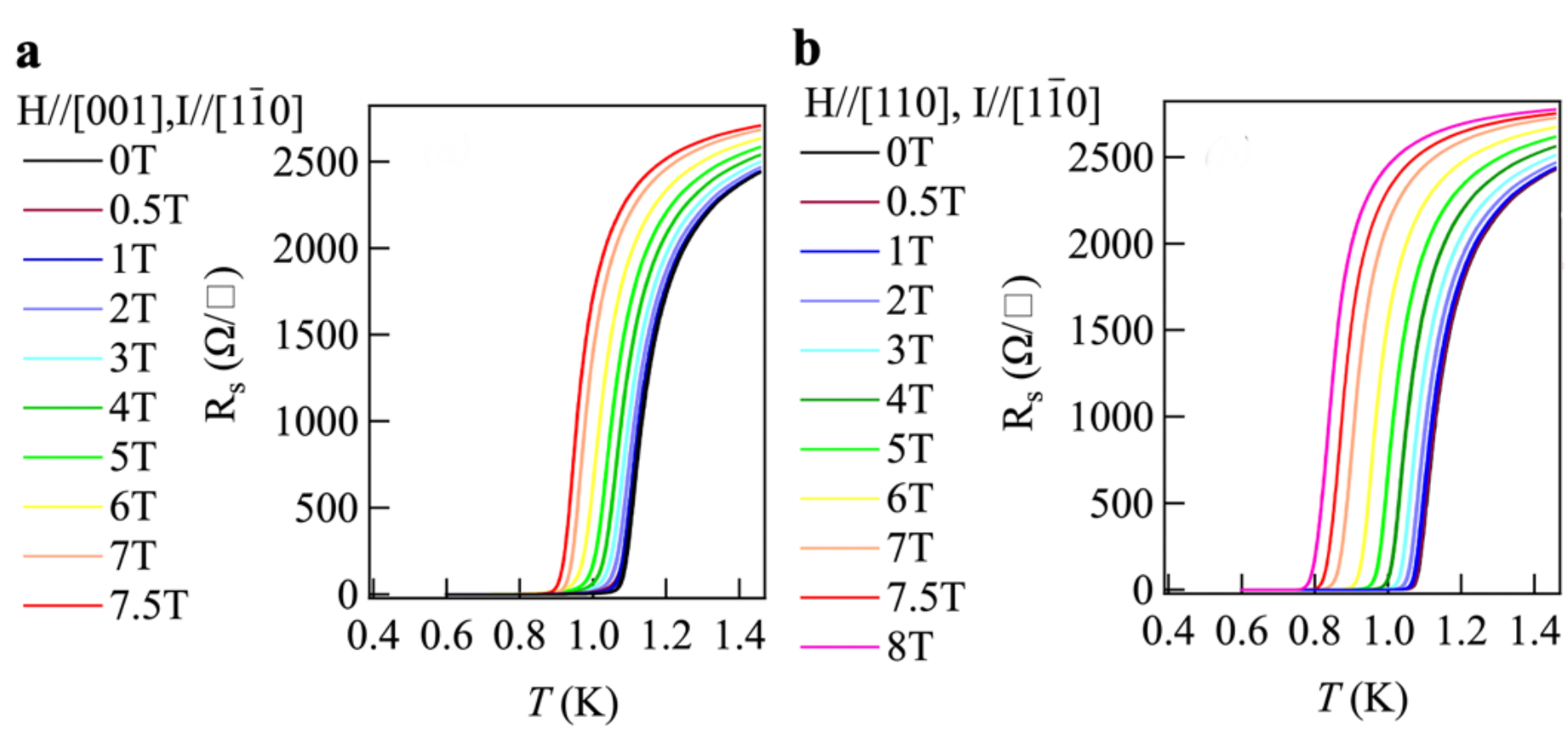

**Figure S8.** $AlO_x$/KTO sample with $T_c$ = 1080 mK Temperature dependence of the resistance for different H//[001]. (b) Temperature dependence of the resistance for different H//[1-10].

## S9: In-plane anisotropy for $AlO_x$/KTO sample with $T_c = 575$ mK

The temperature dependence of the resistance for in-plane fields along [001] and [1-10] are shown in Fig. S9 (a) and (b). The transition temperature is about 575 mK. The critical field vs temperature is extracted from Fig. S9 (a) and (b) and presented in Fig. S9 (c). The critical field is larger for fields along [001] compared to [1-10], displaying the characteristic square-root dependence for a 2D superconductor in a parallel field.

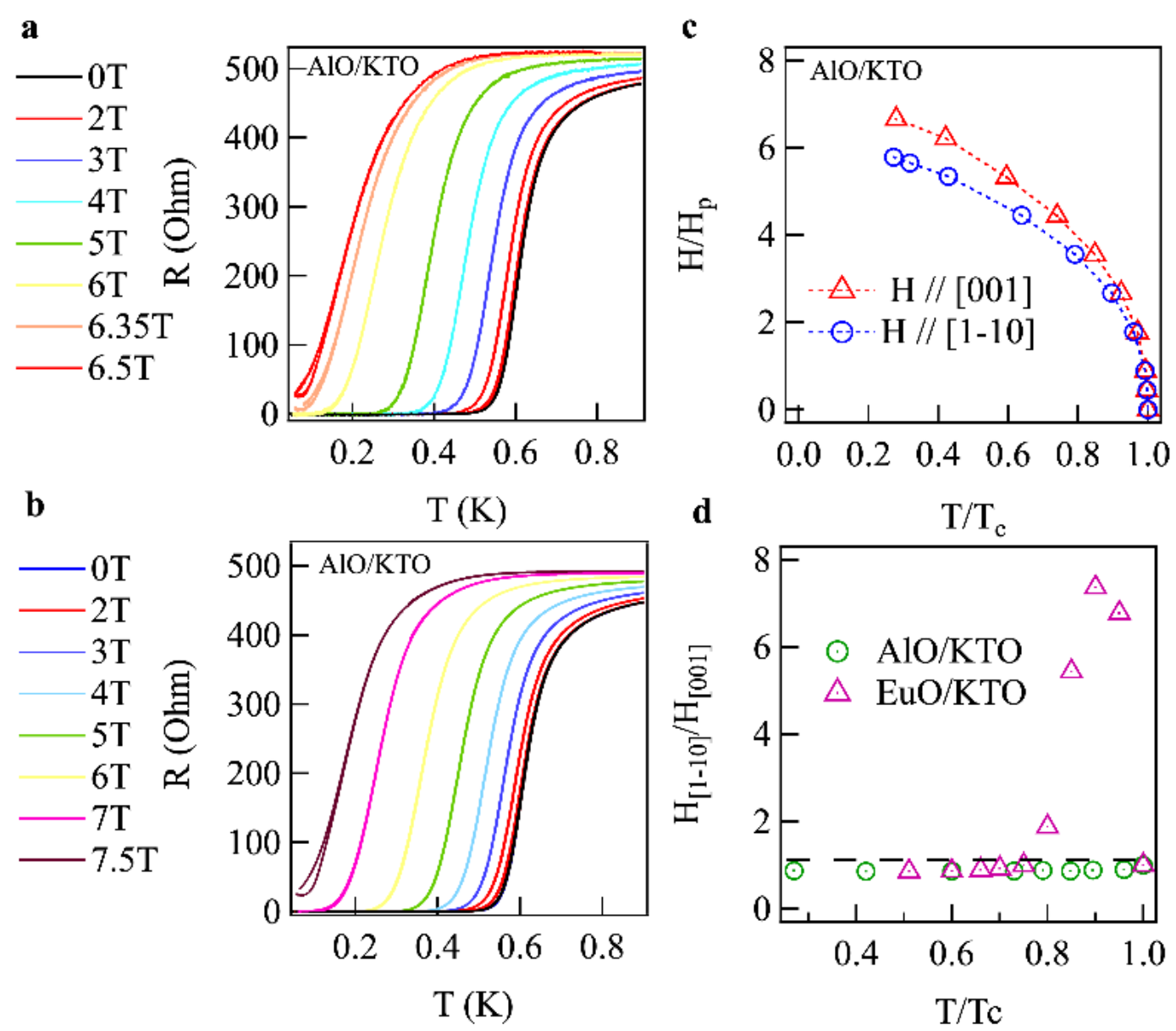

**Figure S9** (a) Temperature dependence of the resistance for different H//[1-10] and (b) for different H//[001]. (c) Critical field vs critical temperature for $AlO_x$/KTO with a $T_c$~ 575 mK for the two field directions. (d) $\frac{H_c\,//\,[1\bar{1}0]}{H_c\,//\,[001]}$ for $AlO_x$/KTO, compared with $EuO_x$/KTO (110) from Fig. 3 in the main text.

## S10: Angular dependence of the resistance for different current directions

Figure S10 presents the angular dependence of the resistance along different field directions for two in-plane current directions near $T_c$ at two values of $H$. The results show that the anisotropic behavior for $EuO_x$/KTO close to $T_c$ is independent of the current direction.

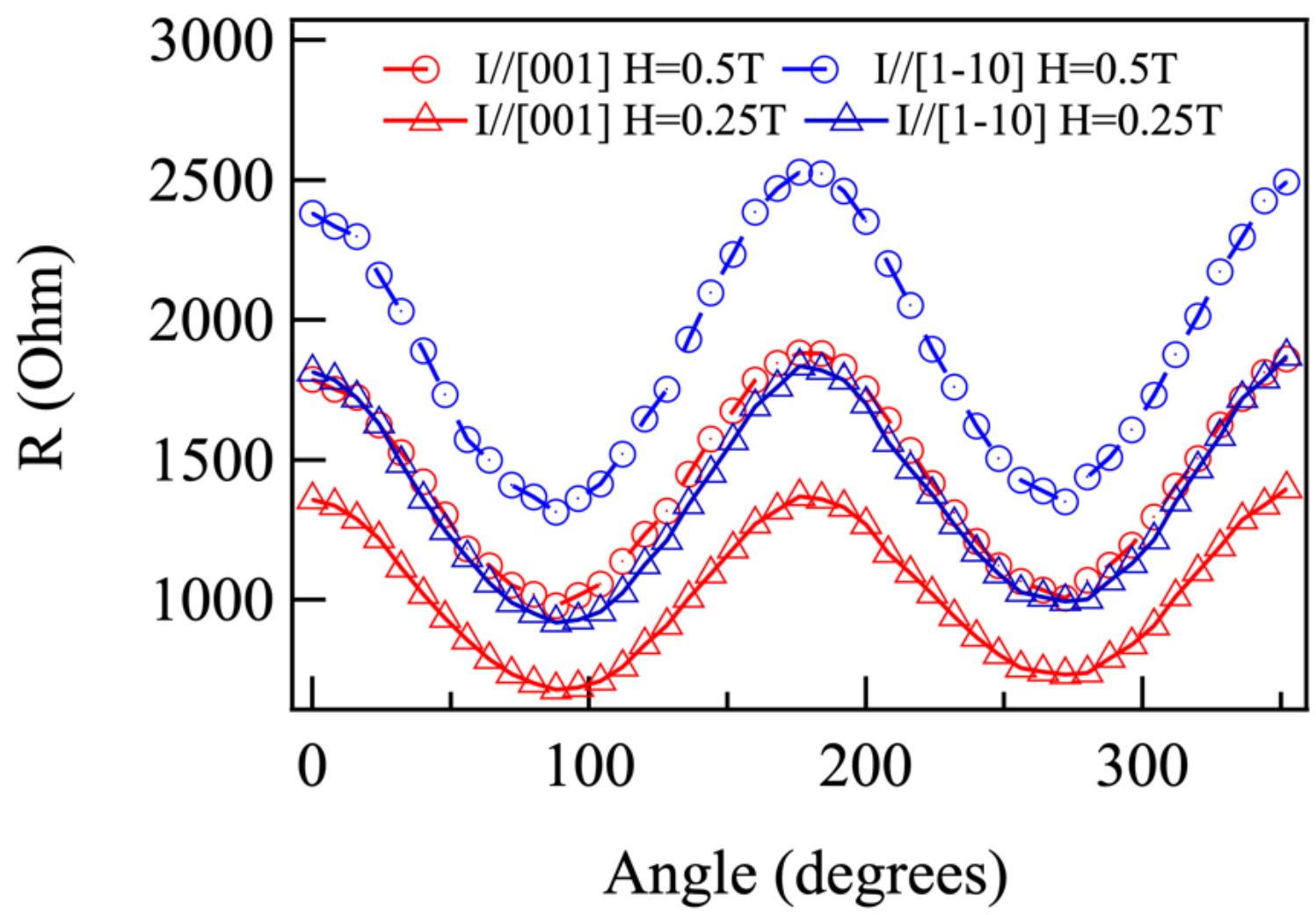

**Figure S*10*** Angular dependence of the resistance at 840 mK with respect to the in-plane field direction for two different values of H for $EuO_x$/KTO(110). Red denotes the current along [001] and blue denotes the current along [1-10].

## S11: Characterizing the $EuO_x$/KTO (110) interface with TEM

The cross-sectional sample was prepared using a focused-ion beam (FIB, model Helios Hydra 5, FEI, USA). The atomic resolution STEM imaging, EDS and EELS, as presented in Figure. 4 and Figure. S11, were conducted on a Spectra 200 TEM (Thermo-Fisher, USA) with a cold field-emission electron source. The microscope operational voltage is at 200 kV. STEM imaging was performed using a HAADF detector with a probe convergence angle of 21.4 mrad and the camera length of 62 mm. The STEM dwell time was 200 ns with 1.29 s per frame. EDS mapping was performed to determine the chemical composition at the interface of EuO/KTO. The EDS spectra were acquired, using a four-quadrant Super-X detector (Thermo-Fisher Scientific) with the acquisition time of 50 μs per spectrum. We collect the EDS spectra in STEM mode for Figure. 4(c-d) over a sample area of 42 x 42 $nm^2$ and for Figure. S11 over an area of 3.4 x 6.7 $nm^2$. The EDS mappings were originally recorded as 60-frame stacks. The electron beam current was kept at 70 pA for both figures. A sample thickness of 40 nm was measured by zero and low loss EELS log ratio. The EELS spectra in Figure. 4(e-f) were collected using a Gatan Image Filter Quantum ER system. EELS spectra were fitted by matrix inversion deconvolution as implemented in Figure. 4(f).

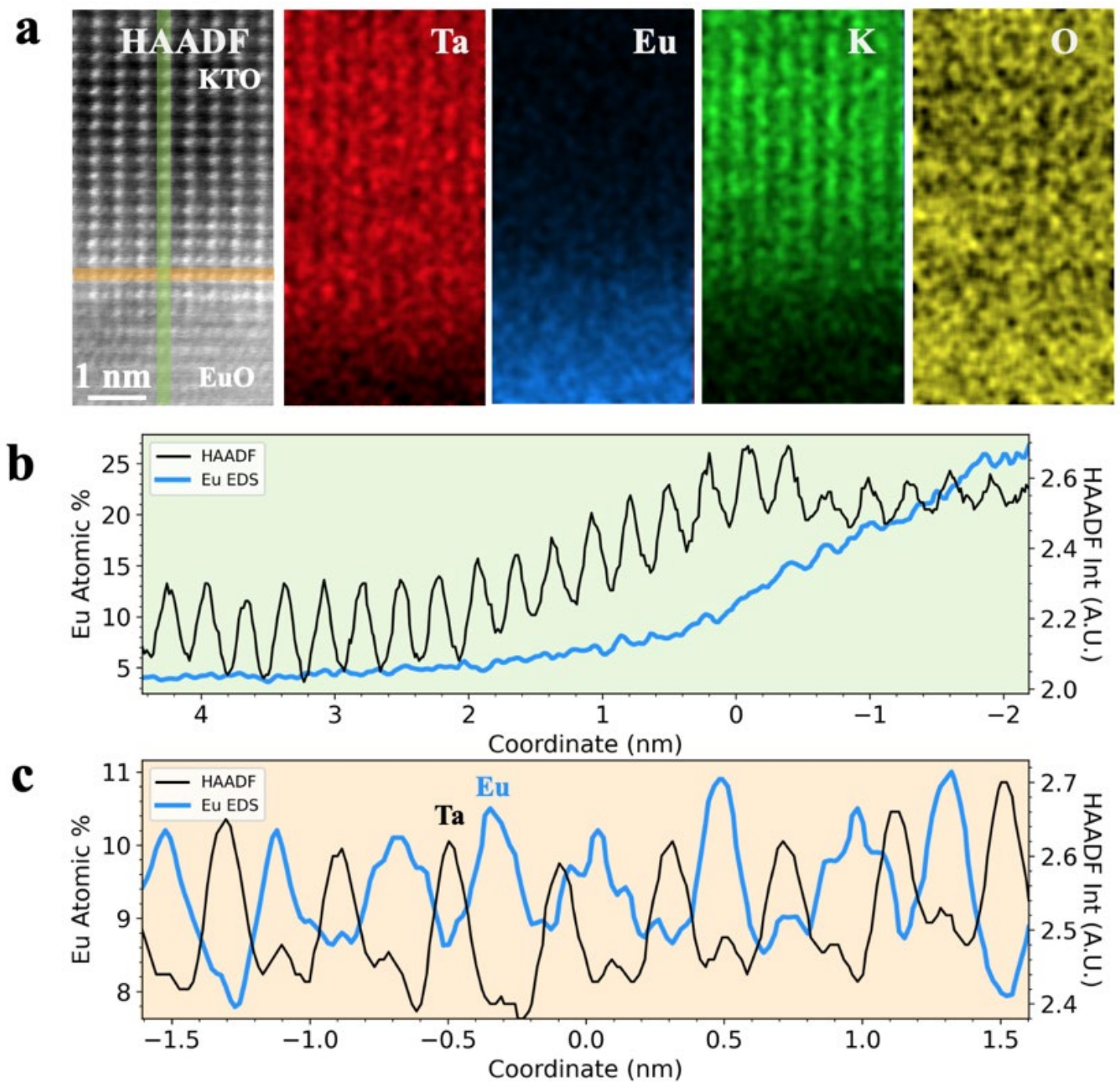

**Figure S*11*** a) High resolution STEM-EDS images of a EuO/KTO[110] interface. b) A vertical line profile of Eu atomic percentage from L and M peak plotted together with the HAADF intensity across the interface marked with a green color line in a). c) A horizontal line profile of Eu atomic percentage from L and M peak and the HAADF intensity at the interface as marked with an orange color line in a). The image size in a) is 3.4 x 6.7 $nm^2$.

## S12: Out-of-plane critical field

The temperature dependence of the resistance under different out-of-plane fields was measured for a $EuO_x$/KTO sample (Fig. S12 (a)). The critical field for out-of-plane fields is significantly smaller compared to the critical field for in-plane fields, as expected given the 2D nature of the superconductivity of the 2DEGs. The critical field vs critical temperature was extracted from Fig. S12(a) and presented in Fig. S12 (b). The Ginzburg-Landau model for a thin film with the field out of-plane gives $H_c = \frac{\Phi_0}{2\pi\xi^2}(1 - \frac{T}{T_c})$. The coherence length $\xi$ is extracted from Fig. S8 (b) and found to be 20.8 nm which is consistent with previous results. The effective thickness of superconductivity state can be extracted from $H_{c,//} = \frac{\Phi_0}{2\pi\xi d}\sqrt{[12(1 - \frac{T}{T_c})]}$. Using the critical field we measured at the lowest temperature in Fig. 3S, we obtain the effective thickness of 6.5 nm (7.6nm) for H//[001] (H//[1-10])

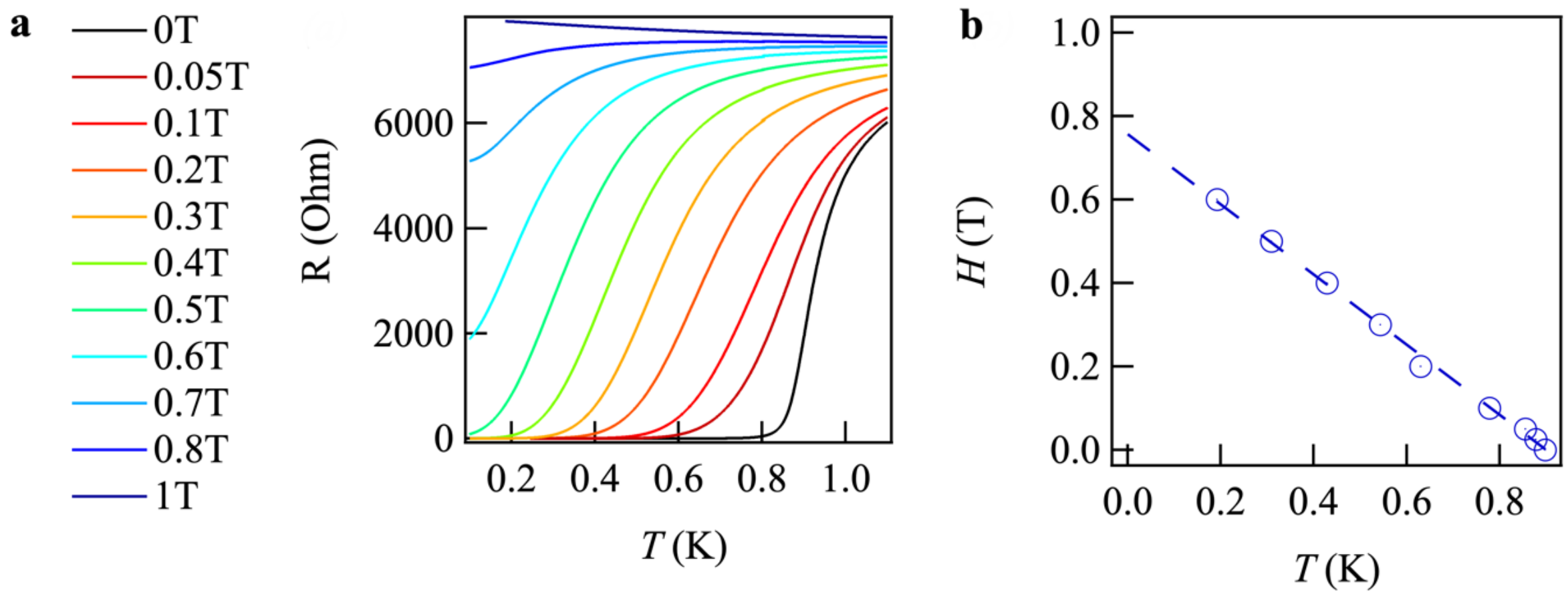

**Figure S12** (a) Temperature dependence of the resistance for different out-of-plane fields for $EuO_x$/KTO(110). (b) Critical field vs critical temperature determined from (a).

## S13: Weak Antilocalization for the field out-of-plane

The magnetoconductance at 2K of a $EuO_x$/KTO(110) and an $AlO_x$/KTO(110) sample are shown when the field is applied out-of-plane in Fig. S13. Both samples show similar weak anti-localization features which is consistent with previous reports[5].

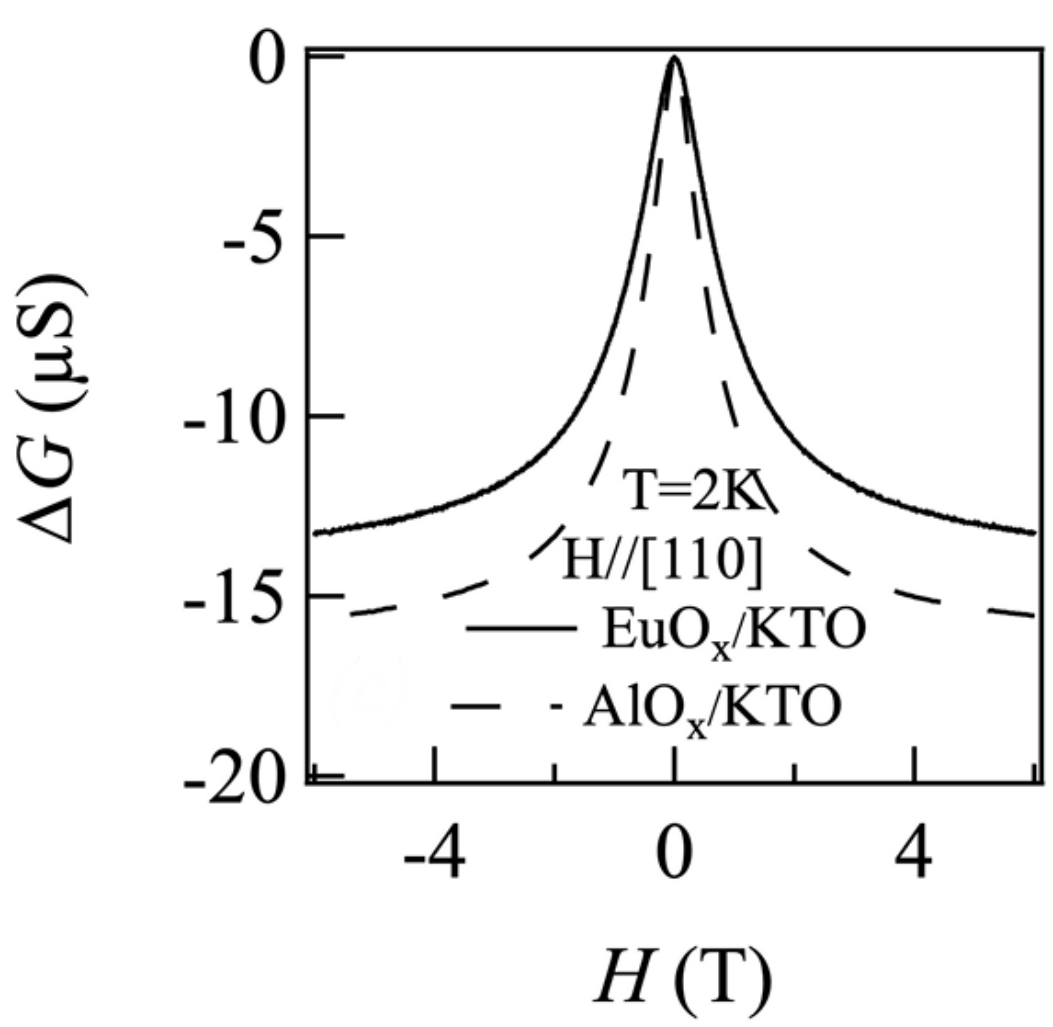

**Figure S13** Field dependence of the magneto-conductance at 2K for $EuO_x$/KTO and $AlO_x$/KTO. The field is applied along the [110] direction which is the out-of-plane direction.

**S14: Theoretical Considerations**

**A. KTO (110) electronic structure**

Due to quantum confinement, the (110) surface has the special property of having low lying *xz*/*yz* states with the *xy* states pushed to higher energy. Considering the *xz*/*yz* subspace, then in the local limit in the presence of spin-orbit, one has spin and orbital moments anti-aligned with equal magnitudes due to Hund's rules, that is $L + 2S = 0$. The net result is that the total moment is quenched. This picture changes once hopping terms are introduced in the Hamiltonian. Given the large band gap in KTO, we can confine to the Ta 5*d* $t_{2g}$ states, noting that the Ta 5*d* $e_g$ states are much higher in energy and so can ignored in an effective low energy model. The dominant $t_{2g}$ hopping, *t*, is the Ta-Ta near-neighbor one (*xz* to *xz* along the *x* and *z* bonds, etc.). For the (110) surface, one of the near-neighbor bond directions is in plane ([001]) and the other two ([100] and [010]) connect planes. This gives rise to a higher Fermi velocity along [001] than along [1-10], resulting in the elongated Fermi surface shown in Fig. 1(b) of the main text. Including just $t$ and the spin-orbit coupling, one finds that the moment texture is completely uniaxial in-plane. That is, the moments are either parallel or antiparallel to the [001] direction since [001] is the quantization axis for *xz*/*yz* moments. The resulting texture around the Fermi surface has the form $-k_y\sigma_x$ where $x$ is along [001] and $y$ along [1-10]. That is, it is "half" the full Rashba form of $k_x\sigma_y - k_y\sigma_x$. Inclusion of other hoppings does not change this picture except when including terms that are off-diagonal in the orbital index. When including this last, a $k_x\sigma_y$ term appears but the net result is that the moment texture is strongly uniaxial, as shown in Fig. 1(b) of the main text. For simulations, we have done both the bilayer and trilayer cases including four hopping terms, $t$, $t'$ (near-neighbor *xz* to *xz* along y, etc.), $t''$ (next-near-neighbor hopping diagonal in the orbital index) and $t'''$ (next-near-neighbor hopping off-diagonal in the orbital index). These hoppings are illustrated in the paper by Xiao *et al*.[6] Values were taken by fitting the bulk band structure of KTO as given in the Materials Project[7]. Results shown were obtained by solving an 18 x 18 secular matrix (three $t_{2g}$ orbitals per Ta site, three layers for the trilayer case, plus spin). The Zeeman term in the Hamiltonian (for a non-zero field) involves both spin and orbital contributions [8].

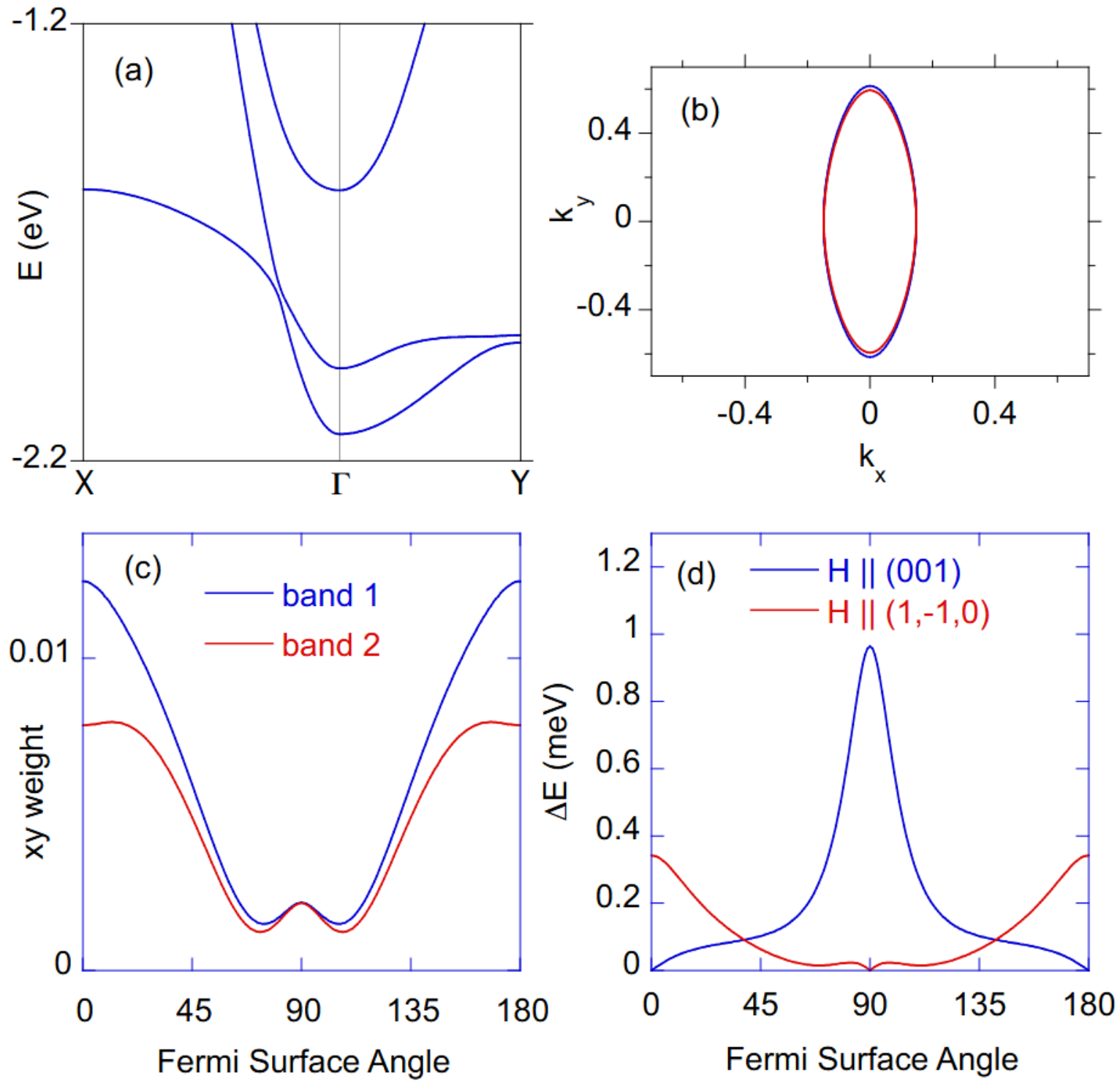

**Figure S14.** (a) Band dispersion for the trilayer case. X is the zone boundary along [001] and Y the zone boundary along [1-10]. The parameters are $t$ = 0.4975 eV; $t'$ = 0.035 eV; $t''$ = 0.09 eV; t''' = 0.0175 eV; and $\lambda_{SO}$ = 0.265 eV. (b) Fermi surface for a density of 6 x $10^{13}$ $cm^{-2}$ (chemical potential $\mu = 0.1321\ eV$). A Rashba interaction ($t_R$ = 2 meV) has been added to the Hamiltonian to lift the Kramers degeneracy, with the Rashba-split surfaces in red and blue. The *x*-axis is along [001] and the *y*-axis along [1-10] with ±1 marking the zone boundaries. (c) Admixture of *xy* states around the Fermi surface. Bands 1 and 2 are the Rashba-split surfaces shown in (b). (d) Zeeman splitting around the Fermi surface for the two in-plane field directions. A Zeeman field of 1 meV is assumed, which for g = 2 would correspond to a 2 meV splitting. Results are shown for band 1.

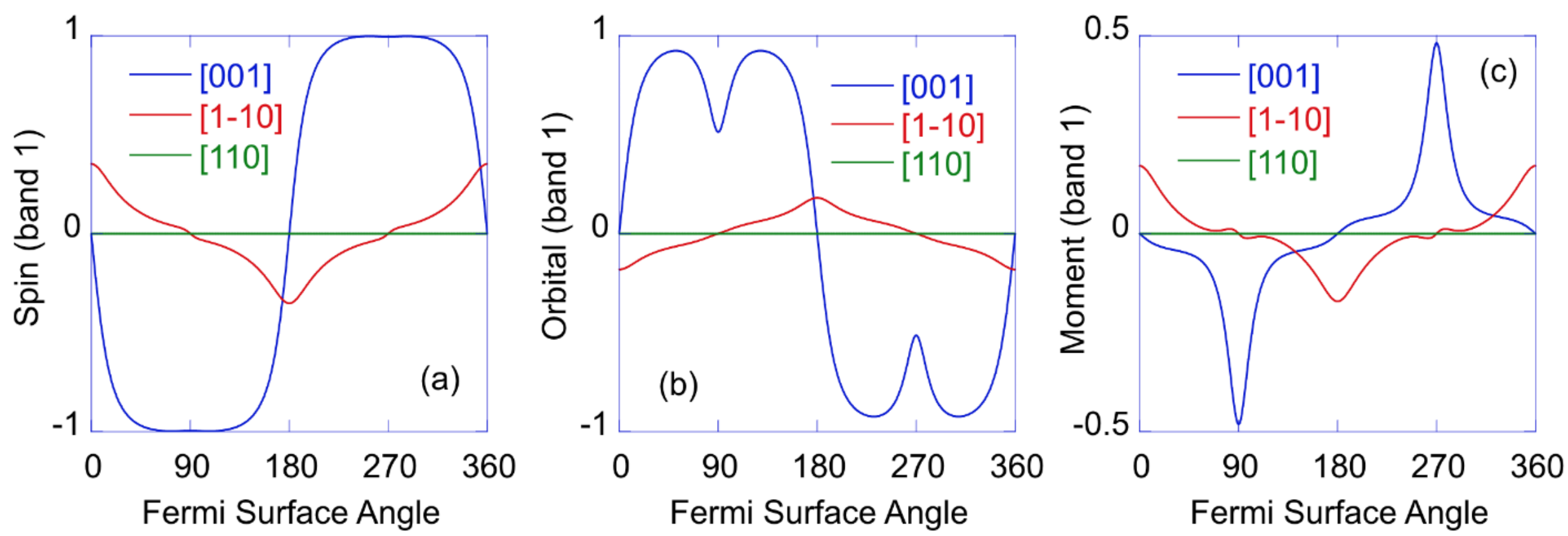

**Figure S15.** (a)-(c) The three components of (a) *S*, (b) *L* and (c) the moment vector (*L*+2*S*) around the band 1 Fermi surface. The maximum allowed would be ±1. Note the band 2 surface has opposite helicity.

In Figure S14, we show results for the trilayer case for (a) the band dispersion, (b) the Fermi surface (corresponding to a carrier density of 6 x $10^{13}$ $cm^{-2}$), (c) the admixture of the *xy* orbitals around the Fermi surface, and (d) the Zeeman splitting around the Fermi surface assuming a Zeeman field of 1 meV. For these plots, a small Rashba interaction of 2 meV has been added to the Hamiltonian to lift the Kramers degeneracy at zero field. Note that the Zeeman splitting for fields along [1-10] follows the *xy* admixture showing its different origin from that along [001]. In Figure S15, the angular dependence of the spin, orbital, and total moment for the three directions is shown around the Fermi surface based on the results from Figure S15. Note that there is no out-of-plane contribution like there is for the (111) surface[8]. In Table 1, we show the expectation values averaged around the Fermi surface associated with $S$ and $M$ for both the bilayer and trilayer cases for the two in-plane field directions, [001] and [1-10], where $M$ is the total moment ($2S + L$, where 2 is the spin $g$-factor). This illustrates the strong anisotropy of the texture with the maximal response being for a field along [001] as expected. The anisotropy is much larger when considering only the spin texture, and the total moment texture is largely quenched due to spin/orbital moment compensation given their opposing signs. When considering these results for the following two subsections, $S$ and $L$ for Ta can in principle couple differently to the Eu ions (that is, not necessarily as $2S + L$ with a typical exchange model involving only $S$).

| *i* | *S* (2L) | *M* (2L) | *S* (3L) | *M* (3L) |
|---|---|---|---|---|
| **[001]** | 0.882 | 0.113 | 0.882 | 0.162 |
| **[1-10]** | 0.165 | 0.108 | 0.173 | 0.077 |
| **[110]** | 0.0 | 0.0 | 0.0 | 0.0 |

**Table 1.** Band 1 averages of the $i^{th}$ component of $S$ and $M$ around the Fermi surface comparing the bilayer simulation (2L) to the trilayer simulation (3L). Here $S^2 = \langle S_i^2 \rangle_{FS}$ and $M^2 = \langle M_i^2 \rangle_{FS}$ (where $M_i = L_i + 2S_i$). The maximum allowed value is 1. Note the large anisotropy for $M$, especially for the trilayer case, and the even larger anisotropy associated with $S$.

## B. In-plane upper critical field anisotropy

The basic formalism we employ is that of Fischer[9] based on WHH[10] theory.The specific formulas employed are for in-plane fields, meaning that the field for the orbital part enters quadratically. Fischer's generalization of WHH is to allow for an extra exchange field in the Zeeman term that accounts for the interaction of the Ta 5$d$ electrons at the Fermi level with the Eu spins. The one generalization we do is to allow the g-factor of the Ta electrons to deviate from 2. The resulting formula for the critical field is:

$$ln\left(\frac{1}{t}\right) = \left(\frac{1}{2} + i\frac{\lambda_{so}}{4\gamma}\right)\psi\left(\frac{1}{2} + \frac{h^2 + (\frac{\lambda_{so}}{2}) + i\gamma}{2t}\right) + \left(\frac{1}{2} - i\frac{\lambda_{so}}{4\gamma}\right)\psi\left(\frac{1}{2} + \frac{h^2 + \frac{\lambda_{so}}{2} - i\gamma}{2t}\right) - \psi\left(\frac{1}{2}\right) \quad \textbf{Eq. 1}$$

where $t$ is the reduced temperature, $T/T_c$ (with $T_c$ the zero field value), $\lambda_{so}$ is the spin-orbit scattering parameter ($\lambda_{so}^{-1} = 3\pi k_B T_c \tau_{so}/2\hbar$ where $\tau_{so}$ is the spin-orbit scattering time), $\psi(x)$ is the digamma function, and

$$\gamma = \sqrt{\{\alpha^2\left[\left(\frac{g}{2}\right)h + h_J\right]^2 - (\frac{\lambda_{so}}{2})^2\}} \quad \textbf{Eq. 2}$$

where $\alpha$ is the Maki parameter (for thin films in a parallel field, its definition differs from the typically quoted bulk/perpendicular field value, $\alpha_B$: $\alpha = 29.058\ \mathrm{sqrt}(\alpha_B/T_c)/d$ where d is the superconducting film thickness in nm, $T_c$ is in Kelvin, and $\alpha_B = \hbar/(2m^*D)$ with m* the effective mass and D the diffusion constant), $h$ is a reduced field, and $h_J$ is the reduced exchange field between the polarized Eu ions and the Ta conduction electrons. For $H$ in Tesla and $T$ in Kelvin, $h = 0.2158\ H / (\alpha\ T_c)$. The exchange field is $h_J = h_{J0}B_J(x)$ where $B_J(x)$ is the Brillouin function for $J = 7/2$ and $x = J\frac{g'\mu_B B}{k_B T}$ with g'=2. The above formula is then optimized over the fit variables using Powell's method [11].

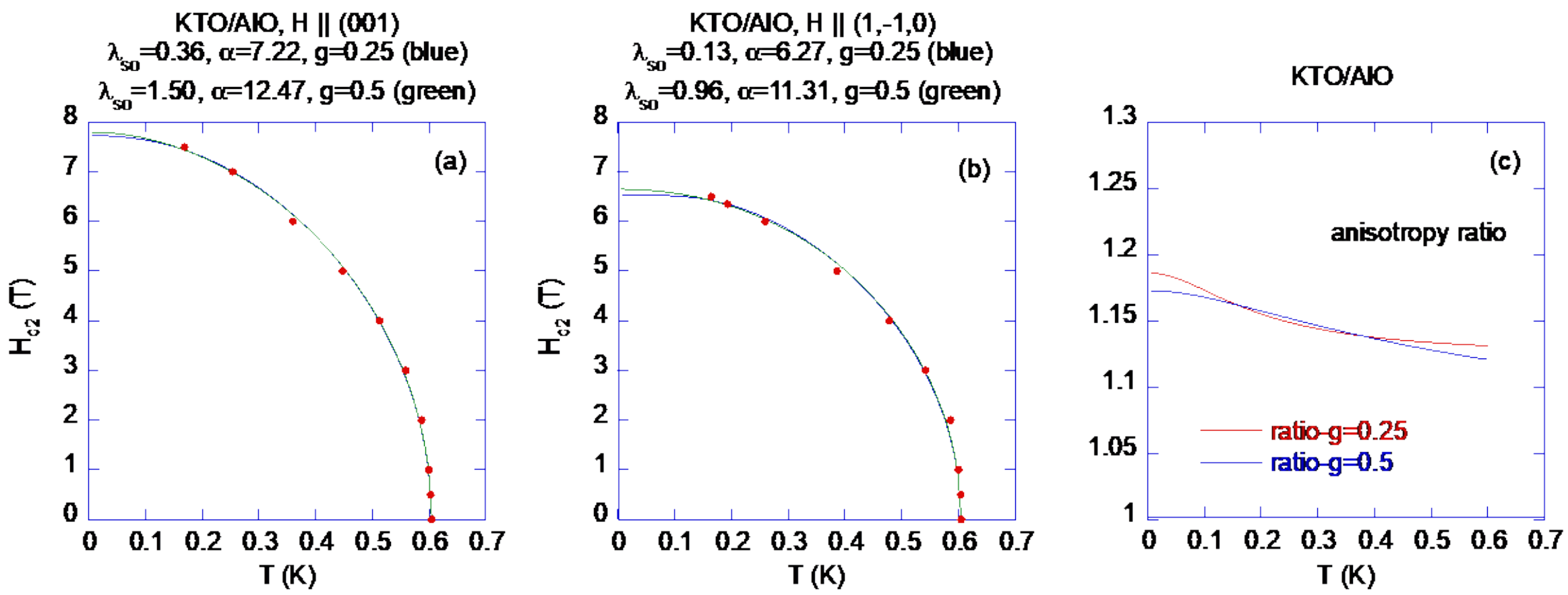

**Figure S16** Fits to the upper critical field vs T for $AlO_x$/KTO for two different values of g for (a) fields along [001] and (b) fields along [1-10]. (c) Anisotropy ratio versus temperature for g=0.25 and g=0.5.

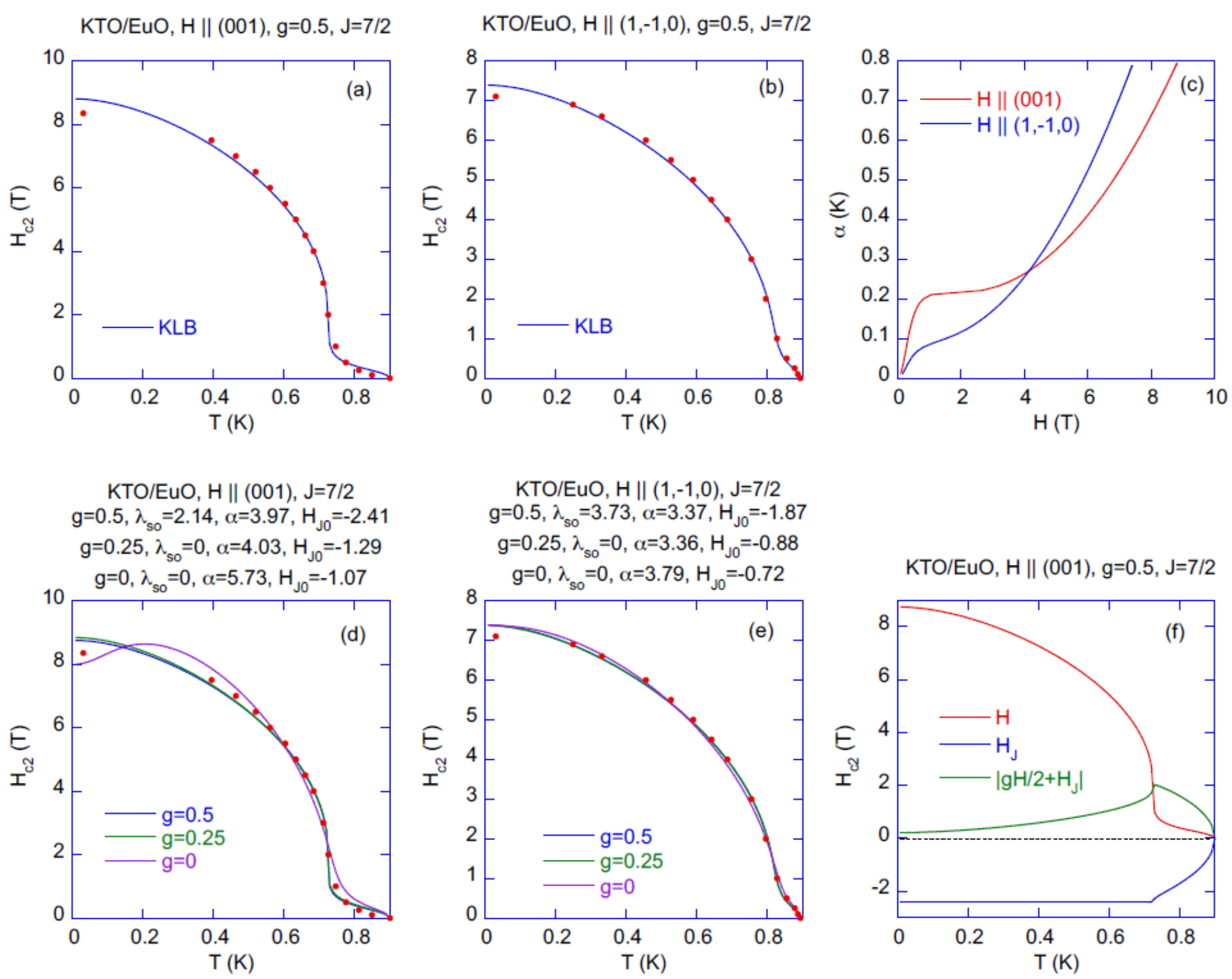

**Figure S17**. Fits to $H_{c2}$ vs T for EuO/KTO (110) using the KLB formalism for (a) fields along [001] and (b) fields along [1-10]. The *g*-factor is assumed to be 0.5 and the impurity $Eu^{2+}$ ion is assumed to have J = 7/2. (c) Plot of the pair-breaking parameter α obtained from the KLB fits (d,e) WHH fits to the same data, for *g*-factors of 0, 0.25 and 0.5, and J = 7/2 for the $Eu^{2+}$ ion (f) Decomposition of $H_{c2}$ for fields along (001) showing the exchange contribution ($H_J$) and the total Zeeman contribution ($gH/2+H_J$).

In Figure S16, we show fits to the critical field for an $AlO_x$/KTO sample with $T_c$ = 600 mK for the two in-plane field directions (where for $AlO_x$, $h_J$ is not included). Given the smooth Ginzburg-Landau nature of the critical field curves, there is no unique determination of parameters given that we have three of them ($\lambda_{so}$, $\alpha$ and $g$). For illustrative purposes, we show results for $g$ = 0.25 and $g$ = 0.5. Note that α differs between the two directions, which is connected to the orbital field anisotropy mentioned in the main text, with $H_{c2}$ scaling with $\alpha$. The anisotropy ratio between the two directions displays a modest temperature dependence. The results are consistent with a finite but suppressed $g$-factor ($g$ acts to suppress $H_{c2}$ at low temperatures) and an anisotropic $\lambda_{SO}$. But the anisotropy of the spin texture is difficult to infer from these measurements, unlike for our observations on EuO/KTO.

In Figure S17, we show fits to the critical field for a EuO/KTO sample with $T_c$ = 900 mK using both KLB and WHH formalisms, where we include the exchange term $h_J$ between $Eu^{2+}$ ions and electrons in the KTO (110) 2DEG. For the KLB fits, we assume $g$ = 0.5. For the WHH fits, we have included fits for $g$ = 0, 0.25 and 0.5. In all fits, $h_J$ is larger for fields along [001] than for [1-10], as expected. When $g$ is non-zero, the sign of $h_J$ matters, and we find that for the best fits, it is negative relative to $h$. This allows for the inflection behavior in $H_c$ (due to rapid rise of the Brillouin function with $H$) to become even more pronounced due to the Jaccarino-Peter effect[12] . This effect is less pronounced for fields along [1-10] reflecting the moment anisotropy shown in Table 1.

## C. Quantum corrections to the normal state magnetoresistance

For parallel fields, the primary contribution to weak antilocalization corrections to the magnetoresistance (MR) is the Zeeman term[13], noting that orbital contributions to the MR exist that will be quadratic in field. Therefore, as outlined in the main text, $H_J$ will enter this Zeeman term as well. This was elucidated earlier for a $NdTiO_3/SrTiO_3$ 2DEG[14]. We assume an exchange field from the Eu ion as in the previous section.

We take the quadratic orbital contribution into consideration and fit the magnetoconductance for both H//[001] and H//[1-10] using the following:

$$\frac{\Delta G}{G}(H_{\parallel}) = m_3 {H_{\parallel}}^2 + m_1 ln(1 + m_2((\frac{g}{2}) H + H_J(x))^2) \qquad \textbf{Eq 3}$$

where the first term on the right corresponds to the orbital correction and the second term to the Zeeman correction. Here, we set $g$=0.5 and use $H_J$ from the g=0.5 fits shown in Fig. S17 with $J$=7/2. The fitted plot is shown in Figure S18 and the fitting parameters are listed in Table. 2. The fits are quite good indicating that the values found from the $H_c$ analysis and the MR analysis are consistent with one another.

| | H//[001] | H//[1-10] |
|---|---|---|
| $m_1$ | -0.0031922 | -0.0040034 |
| $m_2$ | 26.189 | 5.1763 |
| $m_3$ | -0.00091977 | -0.0015494 |

**Table 2.** Fitting parameters for WAL for both *H*//[001] and *H*//[1-10] from Figure S18.

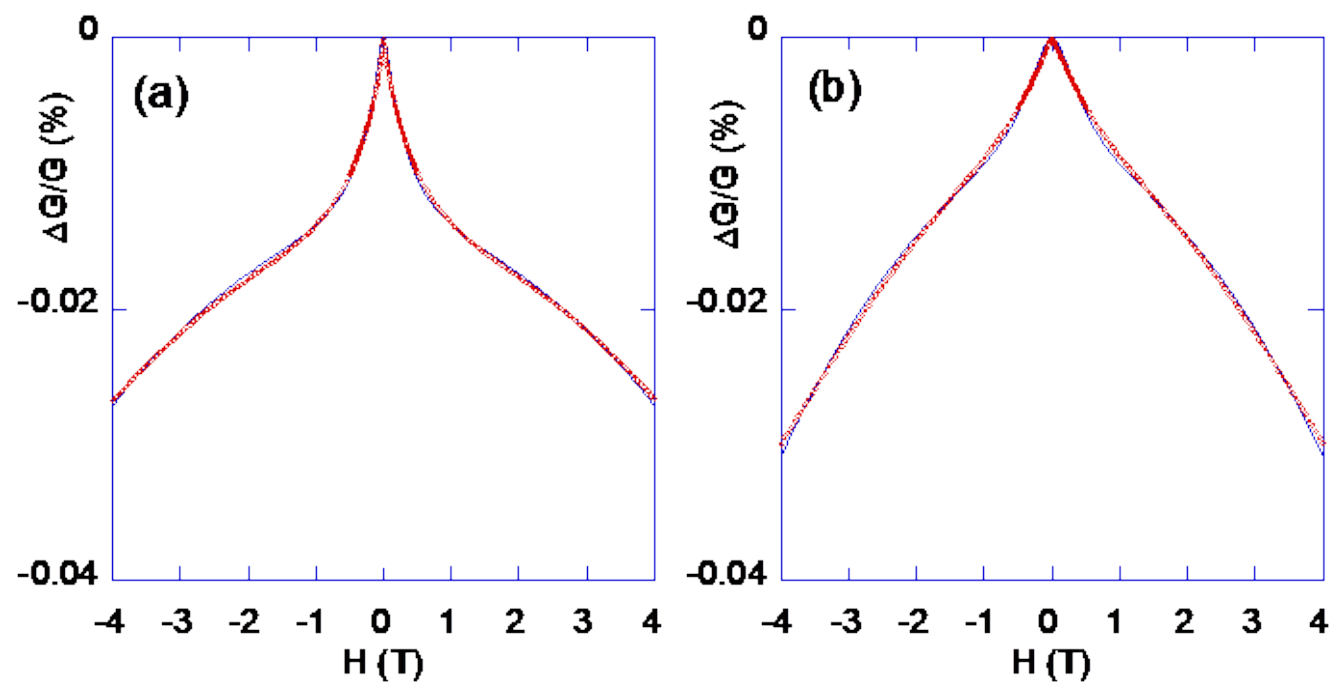

**Figure S18** (a) (b) Fitting of the magnetoconductance at 2K for H//[001] and H//[1-10] respectively for $EuO_x$/KTO(110)